# Physics-informed modularized neural network for advanced building control by deep reinforcement learning

Zixin Jiang, Xuezheng Wang, Bing Dong *

*Department of Mechanical & Aerospace Engineering, Syracuse University, 263 Link Hall, Syracuse, NY 13244, USA*

A B S T R A C T

Physics-informed machine learning (PIML) provides a promising solution for building energy modeling and can be used as a virtual environment to enable reinforcement learning (RL) agents to interact and learn. However, how to integrate physics priors efficiently, evaluate the effectiveness of physics constraints, balance model accuracy and physics consistency, and enable real-world implementation remain open challenges. To address these gaps, this study introduces a Physics-Informed Modularized Neural Network (PI-ModNN), which integrates physics priors through a physics-informed model structure, loss functions, and hard constraints. A new evaluation matrix called "temperature response violation" is developed to quantify the physical consistency of data-driven building dynamic models under varying control inputs and training data sizes. Additionally, a physics prior evaluation framework based on "rule importance" is proposed to quantify the contribution of each individual physical priors, offering guidance on selecting appropriate PIML techniques. The results indicate that incorporating physical priors does not always improve model performance; inappropriate physical priors could decrease model accuracy and consistency. However, hard constraints effectively enforce model consistency. Furthermore, we present a general workflow for developing control-oriented PIML models and integrating them with deep reinforcement learning (DRL). Following this framework, a case study of implementation DRL in an office space for three months demonstrates potential energy savings of 31.4%. Finally, we provide a general guideline for integrating data-driven models with advanced building control through a four-step evaluation framework, paving the way for reliable and scalable implementation of advanced building controls.

## Abbreviation

| | | | |
|---|---|---|---|
| HVAC | Heating, Ventilation, and Air-Conditioning | RL | Reinforcement Learning |
| MPC | Model Predictive Control | DRL | Deep Reinforcement Learning |
| BTM | Behind-the-Meter | VAV | Variable Air Volume |
| DQN | Deep Q-Network | TRPO | Trust-Region Policy Optimization |
| SAC | Soft Actor-Critic | PPO | Proximal Policy Optimization |
| LSTM | Long Short-Term Memory | A3C | Asynchronous Advantage Actor Critic |
| MLP | Multi-Layer Perceptron | AHU | Air Handling Unit |
| BMS | Building Management System | MAE | Mean Absolute Error |
| TRV | Temperature Response Violation | RI | Rule Importance |

## Notation

| | | | |
|---|---|---|---|
| $x$ | State Variable: Zone Air Temperature | $k$ | Time Index: 15 Minute Resolution |
| $u$ | Control Variable: HVAC Thermal Load, Negative for Cooling and Positive for Heating | $c$ | Specific Heat Capacity |
| $w$ | Disturbance Variable: Weather, Occupancy, Time of A Day | $C$ | Thermal Capacity |
| $\boldsymbol{f_{NN_A}}$ | Building Dynamic Network | $y^{T_i}_{meas}$ | Measured Space Air Temperature |
| $\boldsymbol{f_{NN_B}}$ | Control Network | $y^{T_i}_{pred}$ | Predicted Space Air Temperature |
| $\boldsymbol{f_{NN_{E,in}}}$ | Disturbance Network (Internal) | $T_{out}$ | Outdoor Air Temperature |
| $\boldsymbol{f_{NN_{E,ext}}}$ | Disturbance Network (External) | $Q^k_*$ | The average heat flow rate over each time step |
| $\dot{Q}_{sup}$ | Supply Airflow Rate | $\overline{T}_z$ | Setpoint Upper Bound |
| $\dot{Q}_{out}$ | Outdoor Airflow Rate | $\overline{T}_z$ | Setpoint Lower Bound |

(*continued on next page*)

* Corresponding author.
*E-mail address:* bidong@syr.edu (B. Dong).

(*continued*)

| | | | |
|---|---|---|---|
| $x$ | State Variable: Zone Air Temperature | $k$ | Time Index: 15 Minute Resolution |
| $T_{sup}$ | Supply Airflow Temperature | $\ell$ | Reinforcement Learning Reward |

## 1. Introduction

Buildings account for 30% of global final energy use and 27% of energy-related emissions [1], with Heating, Ventilation, and Air-Conditioning (HVAC) systems responsible for over half of this consumption[2]. However, 40% of this energy is wasted due to inappropriate HVAC control, mismatched operation schedules, and other inefficiencies [3]. Therefore, developing advanced HVAC control strategies is crucial for reducing building energy consumption, mitigating global warming, and promoting carbon neutrality.

**Reinforcement learning (RL)** has emerged as a promising approach to advanced building control, where an agent is developed to learn a near-optimal control policy through interaction with the environment, guided by a reward mechanism and iterative trial-and-error processes [4]. Nowadays, with advances in artificial intelligence, RL is increasingly combined with deep learning—known as **deep reinforcement learning (DRL)—**to handle high-dimensional problems using neural networks as function estimator [5]. Compared to model predictive control (MPC), DRL offers higher computational efficiency by offline training before deployment in real buildings and better scalability due to its model-free and data-driven nature [6]. DRL agents can continuously adapt to changing environments through ongoing interaction, requiring minimal human intervention. They provide a promising solution for solving complex building energy optimization problems, such as HVAC control [7], behind-the-meter (BTM) integration [8], supply water temperature control[9], and fan speed control[10].

### 1.1. Deep reinforcement learning for HVAC control

For example, Wei et al. [11] were the first to apply DRL using a deep Q-network (DQN) algorithm for controlling a variable air volume (VAV) HVAC system in 2017. Their proposed method was tested in an EnergyPlus virtual testbed via the BCVTB [12] demonstrated a 20%–70% reduction in energy costs compared to a rule-based baseline. Since then, DRL research in building control has grown rapidly [13]. Biemann et al. [14] compared four actor-critic RL algorithms—soft actor-critic (SAC), proximal policy optimization (PPO), and trust-region policy optimization (TRPO)—in a simulated data center using EnergyPlus. All approaches achieved ∼15% energy savings, with SAC outperforming the others and demonstrating higher data efficiency. Blad et al. [15] trained a DRL agent for underfloor heating system using two black-box environments: a multi-layer perceptron (MLP) and a long short-term memory (LSTM) model. Their method was evaluated on a Dymola virtual testbed and achieved 19.4% cost reductions. Fang et al. [16] developed a DRL agent based on DQN for HVAC system control optimization. The control performance was evaluated using an EnergyPlus-Python co-simulation testbed, demonstrating higher energy efficiency than a rule-based controller while maintaining acceptable temperature violations.

However, the aforementioned studies primarily rely on simulations and only a few studies have evaluated the control performance of DRL through real-world implementations as summarized in Table 1. Such a gap between simulation and real-world implementation may lead to research biases and increase the risk of practical deployment. To address this, a physics-informed dynamic model can provide a reliable and sample-efficient training environment, enhancing real-world applicability.

Note: Where $T_{chilled\text{-}water}$ means chilled water supply temperature; $T_{supply}$ means supply air temperature; $m_{supply}$ means supply air mass flow; $m_{outdoor}$ means outdoor air mass flow; $T_{set}$ means setpoint; $S_{fan}$ means fan speed, $T_{hot\text{-}water}$ means hot water supply temperature; $P_{rad}$ means radiation panel power; FMU (Functional Mock-up Interface), this case study includes four simulated components (building envelope, HVAC system, battery, PV generation); and $P_{bat}$ Means battery charging and discharging power.

### 1.2. Physics-informed building dynamic model

DRL agents learn through trial-and-error and typically require millions of interactions or years of data to develop [25]. However, directly interacting with a real-world system poses significant risks, including potential system failures, safety concerns, and high operational costs. Therefore, in practical applications, a virtual environment serves as the foundation for DRL training, providing a safe and controlled space for policy development.

A reliable virtual environment must combine **fidelity** (accurate representation of real dynamics) and **generalizability** (robustness under unseen conditions). Traditional approaches [26–29] include:

- **White-box models**: Physics-based, interpretable, but labor-intensive.
- **Black-box models**: Data-driven and efficient, but lack interpretability.
- **Gray-box models**: Combine both, but require case-specific calibration.

**Table 1**
Real-world implementations of DRL for HVAC system control and optimization.

| Reference | Building Type | Control Variable | RL Environment | RL Algorithm | Duration | Result |
|---|---|---|---|---|---|---|
| Qiu et al. [17] | Office | $T_{chilled\text{-}water}$ | Real-world chiller plant | Q-learning | 5/1/2021 to 6/30/2021 | from 128 855 to 123 590 kwh |
| Wang and Dong [19] | Office | $T_{supply}$ $m_{supply}$ $m_{outdoor}$ | Physics informed machine leaning | SAC | 9/19/2023 to 2/16/2024 | ∼48% energy saving |
| Lei et al. [20] | Office | $T_{set}$ $S_{fan}$ | Modelica | Branching Dueling Q-network | Four weeks | 14% energy saving and 11% thermal comfort improvement |
| Zhang et al. [21] | Office | $T_{set}$ $T_{hot\text{-}water}$ | EnergyPlus | A3C | 2/6/2018 to 4/24/2018 | ∼16.7% heating reduction |
| Wang et al. [22] | Office | $P_{rad}$ | Physics informed machine leaning | SAC | 22 days | up to 33% energy saving |
| Silvestri et al. [23] | Living Lab | *value* | RC network | SAC | July and August 2023 | 15% to 50% energy savings and 25% comfort improvement |
| Touzani et al. [24] | Office | $T_{supply}$ $M_{supply}$ $P_{bat}$ | FMU | DDPG | Three weeks | 39.6% cost savings and 50% peak reduction |

To leverage the strengths of both physics-based and data-driven approaches while mitigating their respective limitations, the state-of-the-art approach in building dynamic modeling is **physics-informed machine learning (PIML)** [30]. This hybrid methodology integrates physical principles with data-driven learning, enhancing model accuracy, robustness, and generalizability for real-world building control applications. Interested readers can refer to reviews [31,43] for a detailed discussion on the definition, applications, and methodologies of PIML in building performance simulation.

### 1.3. Research gaps and contributions

Despite the growing interest in DRL for advanced building control, several critical research questions remain:

- **How to develop a PIML model effectively and integrated with DRL?**
  Although a few studies have explored PIML for building dynamics, existing models often rely on limited types of physics priors and the integration with DRL remains largely unexplored in real-world settings. More research is needed to explore how PIML can be modularized, generalized, and effectively coupled with DRL to enhance both training and deployment.
- **How can the physical consistency and value of prior knowledge be evaluated?**
  Existing methods for integrating physics into machine learning lack standardized criteria to assess whether trained models adheres to physical laws. Additionally, quantifying the effectiveness and value of each integrated physical rule remains a challenge.
- **What is the general guideline for applying data-driven models to real building control?**
  With increased sensing and data availability, the use of data-driven models in real buildings is growing. However, the reliability of data-driven models remains uncertain. Is there a standardized guideline to validate whether a data-driven model is suitable for advanced building control deployment? Additionally, how can human knowledge be leveraged to refine and improve these models for real-world applications remains open questions.
- **How does DRL perform in real-world experiments?**
  Although DRL has shown promise in simulations, real-world performance remains uncertain. The gap between simulated and actual performance raises concerns about robustness and scalability.

To answer these questions, this study makes the following key contributions:

- **Development of a Physics-Informed Modularized Neural Network**
  A PI-ModNN is developed which incorporates physics-informed model structures, loss functions, and constraints for building dynamic modeling. The proposed mode is then integrated with DRL deployment.
- **Establishment of an Evaluation Framework for Physical Consistency and the Value of Priors**
  A systematic framework is proposed to assess whether models satisfy physical laws and to quantify the contribution of each physics prior to model accuracy and consistency.
- **Guidelines for Data-Driven Control Deployment**
  We propose a practical guideline for evaluating and enhancing data-driven models based on prior knowledge, aimed at improving real-world deployment readiness.
- **DRL Deployment in a Real-World Building**
  A comprehensive DRL control framework incorporating the PI-ModNN is implemented and validated in a real building, serving as a proof-of-concept for scalable, physically consistent DRL deployment in actual building environments.

By addressing these gaps, this study advances the development of PIML for smart building control and its integration with DRL for real-world applications, enhancing both performance and practical feasibility while paving the way for large-scale implementation in the future.

### 1.4. Paper organization

The remainder of this paper is structured as follows: Section 2 covers the detailed methodology, including the overall PI-ModNN-DRL learning framework, case study, data collection, model structure, training process, evaluation of accuracy, physical consistency, and prior knowledge value, as well as DRL integration and experiment setup. Section 3 presents the results, including model performance from different aspects, value of difference prior knowledge and control experiment performance. Section 4 provides a discussion of the findings. Section 5 concludes the study.

## 2. Methodology

This study developed a PI-ModNN to model building thermal dynamics which learn from data while being constrained by physics. It was later used as a virtual environment for offline training of a DRL agent. The trained agent was then deployed in a student office to evaluate its real-world control performance. The overall methodology was presented in Fig. 1.

First, a database was developed to collect real-time measured data, including HVAC operation data, indoor environmental conditions, weather data, and occupancy information. Using this dataset, the PI-ModNN was trained to accurately capture the building's thermal behavior.

Next, we proposed a model evaluation framework. The first step involves a standard accuracy evaluation to ensure the model fits the data accurately. The second step evaluates physical consistency both quantitatively and qualitatively—this includes checking the sign of model derivatives and conducting sanity checks by applying different control inputs and identifying abnormal temperature responses. In the third step, if the model fails the physical evaluation, we quantify the contribution of different physics-based priors and select appropriate ones to improve the model's consistency. This framework not only ensures predictive accuracy but also enhances physical consistency, which is critical for safe and reliable control in real-world applications.

After validation, the PI-ModNN was used as the training environment for a DRL agent. The agent interacted with this virtual environment to learn an optimal control policy. The physical consistency of PI-ModNN enabled realistic and reliable training responses, improving sample efficiency and safety. Once the agent achieved satisfactory performance, it was deployed in the actual building via the Building Management System (BMS) to evaluate real-world control effectiveness.

### 2.1. Case study and data collection

Fig. 2 shows the testbed configuration and sensor layout. Additional details are provided in *Supplementary Material: Case Study and Data Collection*. Data was collected at 15-minute intervals from September 2022 to March 2025 (ongoing), with the data distribution shown in *Figure S1*. The lab was under setup until March 2023. Human subject testing was conducted from March to July 2023, followed by optimal control testing from July 2023 until now. Notably, after April 2024, all students moved out of the office, and heat lamps were used to replace internal heat gain. The internal heat gain per occupant was estimated at 80 W [32]; to account for additional plug load contributions, we used heat lamps with a power rating of 150 W to represent each occupant, starting in January 2025. These changes affect the building's thermal response and are considered in training data selection, which is discussed in a later section.

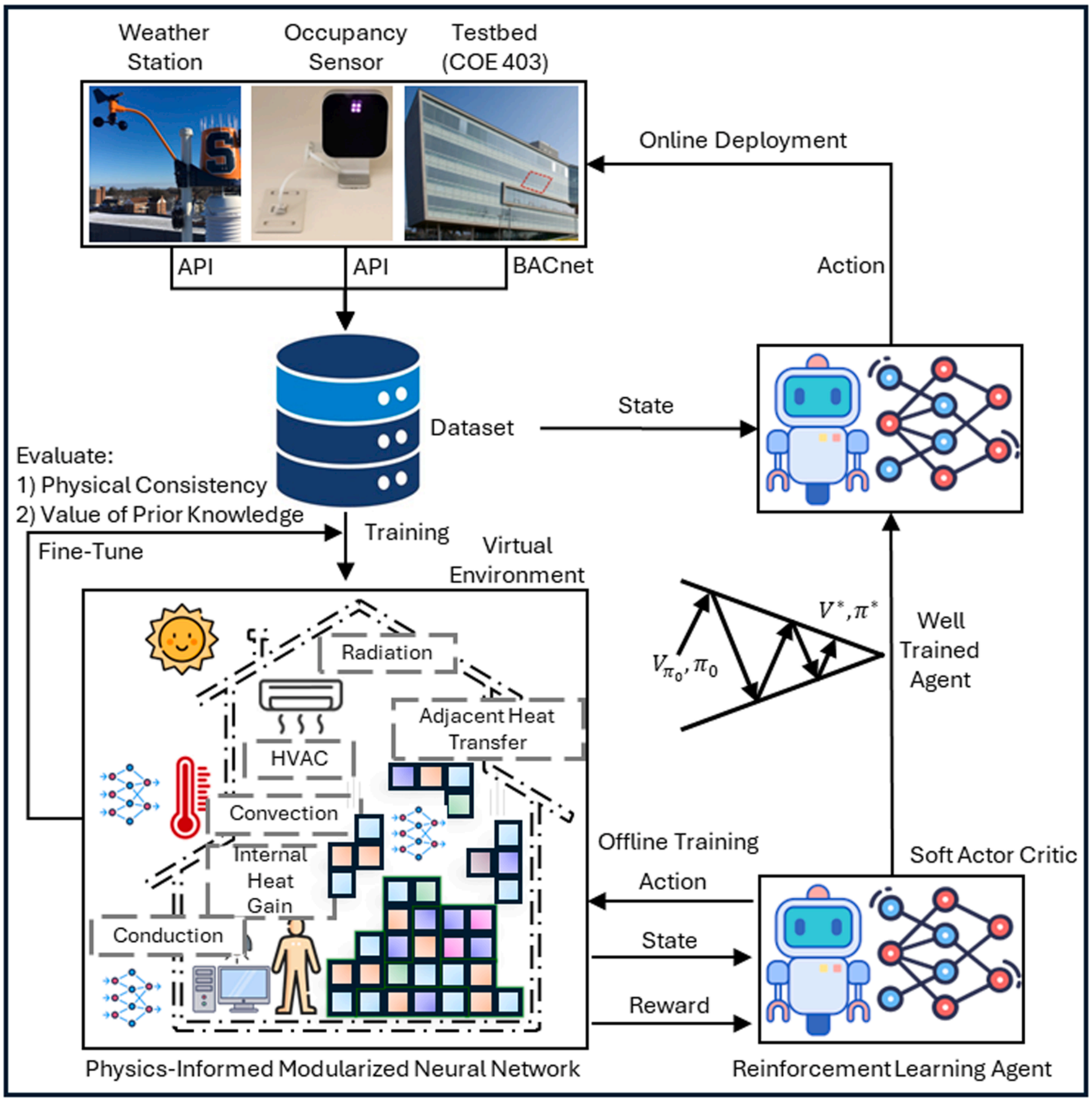

**Fig. 1.** Overall diagram of PI-ModNN-DRL learning framework.

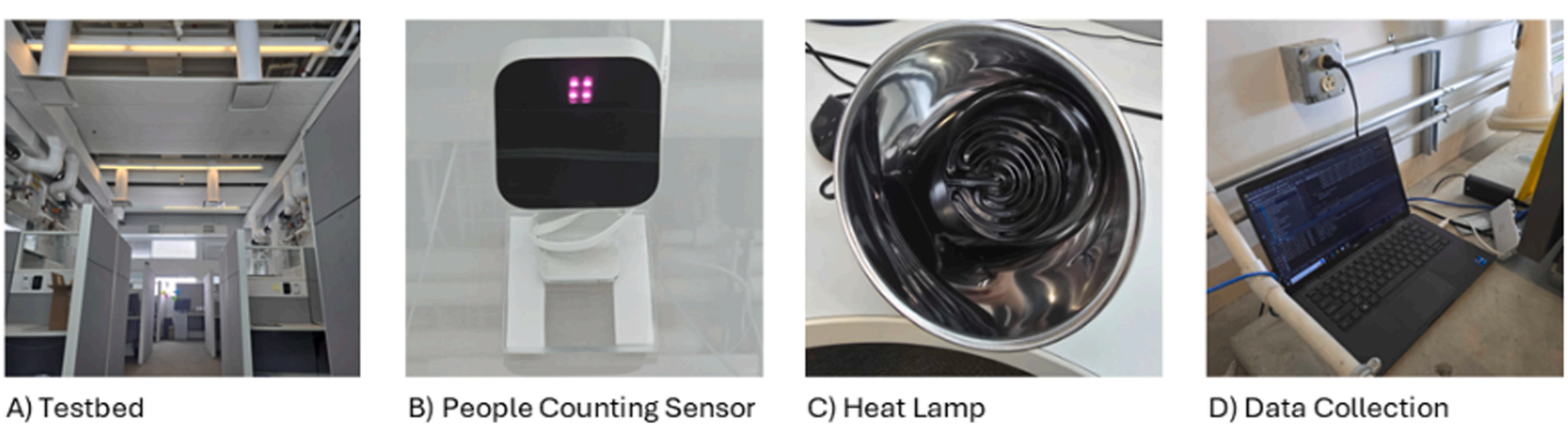

**Fig. 2.** Testbed setup, including the people-counting sensor, heat lamp, and data collection devices.

### 2.2. Physics-informed modularized neural network

#### 2.2.1. Model development

The thermal dynamic model used in this study builds upon our previous work [33], where a modularized neural network was developed, with each module estimating a distinct heat transfer term. We updated the model configuration, and the key modifications are summarized below:

**1) Physics-informed model structure**

The detailed model structure, depicted in Fig. 3, is formulated as a state-space-informed time stepper model [34] for multiple steps ahead prediction. It takes ambient air temperature, solar radiation, time features (sine and cosine), HVAC thermal load from time step $k - encoder$ to $k + decoder - 1$, and space air temperature from $k - encoder$ to $k$ as inputs, to predict the space air temperature from time step $k+1$ to $k + decoder$.

We begin by discretizing the continuous heat balance equation into the form shown in Eq 1,

$$C\frac{dx}{dt} = \dot{Q}_{ext} + \dot{Q}_{int} + \dot{Q}_{hvac} \rightarrow x^{k+1} = x^k + \frac{\Delta t}{C}\left(Q_{ext}^k + Q_{int}^k + Q_{hvac}^k\right) \tag{1}$$

Where the $x$ is the state variable representing the space air temperature, $k$ is the timestep, and $C$ denotes the thermal capacity of the space, calculated as $C = c_{air} \cdot \rho_{air} \cdot V_{air}$, with $c_{air}$, $\rho_{air}$, $V_{air}$ representing the specific heat, density, and volume of air, respectively. $\Delta t$ is the discretized time resolution and $Q^k$ denotes the average heat flow rate over each time step.

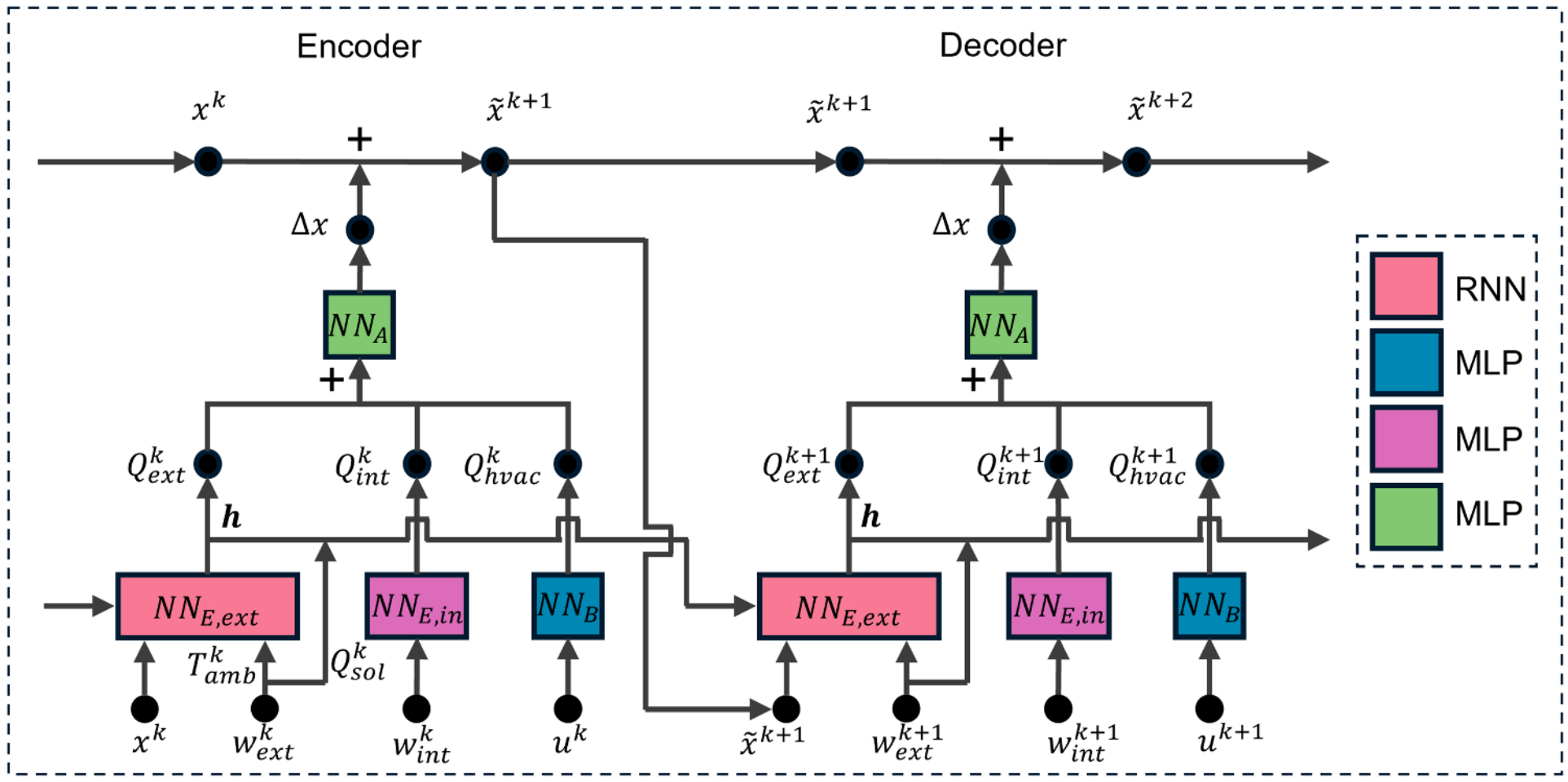

**Fig. 3.** Model structure of proposed PI-ModNN.

The learning objective is to estimate the heat flow components: $Q^k_{hvac}$ from the HVAC system, $Q^k_{int}$ from internal heat gains (e.g., occupants, appliances, lighting), and $Q^k_{ext}$ from external sources and adjacent zones (in multizone scenarios), along with the unknown thermal capacity of the space. The model structure is represented in Eq 2

$$x^{k+1} = x^k + \boldsymbol{f}_{\boldsymbol{NN_A}}\left(\boldsymbol{f}_{\boldsymbol{NN_{E,ext}}}(x^k,\ w^k_{ext}) + \boldsymbol{f}_{\boldsymbol{NN_{E,in}}}(w^k_{int}) + \boldsymbol{f}_{\boldsymbol{NN_B}}(u^k)\right) \tag{2}$$

Where $\boldsymbol{f}_{\boldsymbol{NN_{E,ext}}}$ and $\boldsymbol{f}_{\boldsymbol{NN_{E,in}}}$ are neural networks modeling the external and internal heat gains, respectively; $\boldsymbol{f}_{\boldsymbol{NN_B}}$ models the HVAC heat input; and $\boldsymbol{f}_{\boldsymbol{NN_A}}$ learns the thermal dynamics and acts as an integrator over time. A residual connection is used to update the current state by adding the learned heat input to the previous state. A more detailed discussion of each module is provided in the discussion section.

**2) Physics-informed loss function**

As discussed in prior studies [43], physics-informed loss functions can be broadly classified into three types:

- **Governing equation-based loss**, derived directly from physical governing equations;
- **Knowledge-guided loss**, incorporating empirical rules or expert expertise;
- **Surrogate loss**, using physics-based surrogate models to guide learning in the latent space.

This study focuses on a **knowledge-guided loss** to illustrate a generalizable framework for integrating physics priors into data-driven models. Rather than prescribing a specific formulation, we aim to demonstrate how such priors can be applied and evaluated—recognizing that not all contribute positively due to potential user-introduced bias.

A common challenge in time-series prediction is balancing the bias-variance tradeoff [37,38]. A complex model may successfully capture the overall temperature trend but over-smoothing and fail to reproduce short-term fluctuations driven by changes in HVAC power. To address this, we introduce a short-term variation loss, encouraging the model to learn both the absolute temperature and its dynamic sensitivity to control inputs. The formulation is shown in Eq 3 to Eq 5:

$$l_{accuracy} = \frac{1}{N}\sum_{i=1}^{N}\left(\boldsymbol{y}^{T_i}_{meas} - \boldsymbol{y}^{T_i}_{pred}\right)^2 \tag{3}$$

$$l_{fluctuation} = \frac{1}{N}\sum_{i=1}^{N-1}\left|\left(\boldsymbol{y}^{T_{i+1}}_{meas} - \boldsymbol{y}^{T_i}_{meas}\right) - \left(\boldsymbol{y}^{T_{i+1}}_{pred} - \boldsymbol{y}^{T_i}_{pred}\right)\right| \tag{4}$$

$$l_{total} = l_{accuracy} + \alpha \cdot l_{fluctuation} \tag{5}$$

Here, $l$ denotes the loss function. The terms accuracy and fluctuation represent the accuracy loss and the temperature change per timestep loss, respectively. Total refers to the overall loss, and $\alpha$ is the balancing weight, we use 0.3 in our default setting. $N$ is the number of sampled trajectories, while $\boldsymbol{y}^{T_i}_{meas}$ and $\boldsymbol{y}^{T_i}_{pred}$ denote the measured and predicted space air temperature trajectories at the $i^{th}$ timestep, respectively. A sensitivity analysis of different loss mixing ratios is provided in *Figure A1*. There are many alternative strategies for designing such loss terms. One example is "DIstortion Loss including shApe and TimE" [39], which penalizes errors more strongly when the predicted time series deviates in shape or is temporally misaligned. Other regularization terms can be introduced as well to address non-physical behaviors, such as unrealistic temperature spikes or oscillations.

**3) Physics-informed model constraints**

To ensure physical consistency, models should produce responses aligned with physical laws—e.g., space temperature should decrease with increased cooling and increase with heating. In other words, the gradient of the state variable (e.g., zone air temperature $\boldsymbol{x}_t$) with respect to model inputs—such as disturbances $\boldsymbol{w}_{t-1}$ or control inputs $\boldsymbol{u}_{t-1}$ should be positive no matter for cooling or heating scenarios. A detailed explanation can be found in [40,41,33].

In this case study, the gradient is formulated in Eq 6 using the chain rule:

$$\frac{\partial \boldsymbol{x}_t}{\partial \boldsymbol{u}_{t-1}} = \frac{\partial \boldsymbol{x}_t}{\boldsymbol{f}_{\boldsymbol{NN_A}}}\cdot\frac{\boldsymbol{f}_{\boldsymbol{NN_A}}}{\boldsymbol{f}_{\boldsymbol{NN_B}}}\cdot\frac{\boldsymbol{f}_{\boldsymbol{NN_B}}}{\partial \boldsymbol{u}_{t-1}} > 0 \tag{6}$$

Here, $\boldsymbol{f}_{\boldsymbol{NN_A}}$ and $\boldsymbol{f}_{\boldsymbol{NN_B}}$ represent fully connected neural networks, where each layer is defined as $y = \omega x + b$. Enforcing $\omega > 0$ ensures non-negative gradients. Since they use ReLU activation functions, which inherently produce non-negative outputs. For external module $\boldsymbol{f}_{\boldsymbol{NN_{E,ext}}}$, which is an RNN, the positivity constraint can be directly imposed by restricting the weights to be positive, as demonstrated in study [41].

### *2.2.2. Model training*

Following model development, the dataset was segmented into clean, contiguous sequences without missing values. To improve performance and generalization, early stopping and a mixed training strategy were applied. Detailed training procedures, including a comparison of loss decay with and without mixed training (*Figure A2*) and hyperparameter tuning (*Table S1*), are provided in the *Supplementary Material: Model Training.*

#### 2.2.3. Model evaluation

*2.2.3.1. Performance metrics for accuracy.* The model accuracy is evaluated based on the commonly used mean absolute error (MAE), calculated by Eq 7.

$$MAE = \frac{1}{N}\sum_{i=1}^{N}\left|y_{meas}^{T_i} - y_{pred}^{T_i}\right| \tag{7}$$

*2.2.3.2. Performance metrics for physical consistency.* Beyond accuracy, physical consistency is critical for control applications. A reliable model must produce reasonable responses to changes in control inputs—enabling the DRL agent to explore optimal policies effectively. We propose two evaluation methods:

**1) Qualitative Evaluation**

We evaluate the partial derivative of space air temperature $\boldsymbol{x}$ with respect to the control input $\boldsymbol{u}$ to verify whether the sign of the control gain is positive, as discussed earlier. This can be obtained by automatic differentiation function, which is available in most deep learning packages, such as Pytorch and TensorFlow.

**2) Quantitative Evaluation**

While sign correctness is necessary, it does not guarantee appropriate magnitude or direction of response. We introduced a sanity check by applying five HVAC power levels: −4 kW, −2 kW, 0 kW, +2 kW, and +4 kW. The predicted temperature should decrease with increasing cooling power, in line with energy conservation. Deviations from this expected trend are accumulated as violations (Temperature Response Violation: TRV) and quantified using Eq 8 and Eq 9:

$$TRV^{+} = sum\left(\min\left(T_{check} - T_{pred}\right), 0\right) \tag{(8)}$$

$$TRV^{-} = sum\left(\min\left(T_{pred} - T_{check}\right), 0\right) \tag{9}$$

Where $T_{check}$ represents the space air temperature under a modified sanity check control input, $T_{pred}$ is the model-predicted space air temperature under the original control input and $TRV^{+}$and $TRV^{-}$ capture over- and under-estimation, respectively.

*2.2.3.3. Value of prior knowledge.* After evaluating accuracy and physical consistency, the next question is: What is the contribution of each physical rule? Addressing this helps determine whether specific rules should be implemented and provides clear guidance for model adjustment

To assess the effectiveness of prior knowledge, we use the concept of **Rule Importance (RI)**[42], where the contribution of each physical rule is measured by its impact on model performance in terms of accuracy and consistency. The value of prior knowledge is defined as

$$RI_s(i) = \log_{10}(f(s) + \varepsilon) - \log_{10}(f(s \cup \{i\}) + \varepsilon) \tag{10}$$

Where $s$ is the rules applied in baseline, $i$ is the rules that we aimed to evaluate, $\varepsilon$ is a small constant, e.g., $1e^{-6}$ that prevents log(0) errors, $f$ is the performance index such as $MAE$ or $TRV$ in this study, $s \cup \{i\}$ is the new set by adding rule $i$.

#### 2.2.4. Model comparison

To evaluate the effectiveness of the proposed PI-ModNN and the value of different physics priors, we compare five model configurations:

- **LSTM**: Purely data-driven baseline model.
- **PI-ModNN|LC**: Modularized neural network without physics-informed loss or constraints.
- **PI-ModNN|L**: With constraints but without physics-informed loss.
- **PI-ModNN|C**: With loss terms but without physics-informed constraints.
- **PI-ModNN**: Fully physics-informed modularized neural network.

Each model is evaluated on both accuracy and physical consistency. Given the dataset spans nearly three years, we conduct experiments using training data of varying lengths: 7, 30, 90, 180, and 300 days. Each model is trained 30 times per setting and evaluated on August 2024 using a 96-step prediction horizon (24 hours ahead). Performance is calculated using a rolling window, resulting in 2,976 predictions (96 steps × 31 days) per model per experiment for robust comparison.

### 2.3. Deep reinforcement learning development

We adopt the SAC algorithm[18], a state-of-the-art model-free DRL method that maximizes both expected reward and policy entropy. Compared to alternatives such as DDPG and TD3, SAC provides superior sample efficiency, training stability, and robustness in continuous control tasks. It follows an actor–critic framework enhanced by entropy regularization, enabling effective exploration–exploitation balance.

In this study, we use the similar DRL configuration as our previous work [19]. First, the PI-ModNN is trained using historical data to model the building thermal dynamics. The PI-ModNN takes control inputs (e.g., HVAC power) and disturbance inputs (e.g., ambient air temperature, solar radiation) to predict future space air temperature. After training, the PI-ModNN serves as a surrogate environment for the SAC agent to learn the optimal control policy.

During training, the SAC agent samples a control action (i.e., one-step HVAC power) from its stochastic policy network based on current observations, including indoor air temperature, ambient air temperature, solar radiation, occupancy, and time features. The proposed control action, along with the disturbance inputs, is passed to the pre-trained PI-ModNN to predict the next space air temperature.

The reward is computed using Equations *Eq S1* to *Eq S6* (*Supplementary Material: DRL Development*), and the full transition tuple (state, action, reward, next state) is stored in a replay buffer. A hybrid offline-to-online training strategy is used to update the policy, value, and Q networks via mini-batches from this buffer.

We evaluate three DRL training conditions: using only simulated data, only measured data, and a combination of both—each with both LSTM- and PI-ModNN-based environments, as shown in *Figure A3*. Training with PI-ModNN shows superior learning performance, while measured-only training fails to converge, and simulated-only training yields the highest reward. Further details on the reward function design, hybrid training strategy, DRL hyperparameters, and training strategy are provided in the *Supplementary Material: DRL Development.*

### 2.4. Experiment setup

Detailed experimental setup is summarized in the *Supplementary Material: Experiment Setup*, which includes both the baseline and the DRL control logic. At the equipment level, both baseline and DRL use identical PI controllers. At the system level, The baseline control logic follows the occupancy-based temperature and ventilation control, as used in our previous studies[44,19], and is aligned with the recommendations of ASHRAE Guideline 36-2021: High-Performance Sequences of Operation for HVAC Systems. Temperature setpoints, supply airflow, outdoor airflow and supply air temperature are summarized in *Table S4*, and ventilation rates are calculated using *Eq S7*. And the detailed schedule of the experiment is shown in *Figure S2*.

## 3. Result

### 3.1. Model accuracy evaluation

*Figure A4* show two-week predictions from the LSTM and PI-ModNN models, both of which closely track the measured data. The LSTM and PI-ModNN models achieved MAEs of 0.32°C and 0.30°C, respectively, demonstrating strong predictive accuracy. To further evaluate performance, we analyzed the effect of training data size on model accuracy (Fig. 4A and *Table A1*). The LSTM model performed well across all training durations (7 to 300 days). This may be due to the stable indoor conditions of the small testbed, allowing LSTM to effectively capture average trends. However, model performance slightly declines when using 90-day and 180-day training data. This may be because of the seasonal inconsistency between training and testing periods: the testing period is in August (cooling season), whereas the 90-day and 180-day training datasets include both heating and cooling seasons. The inclusion of mixed-season data, without any physics-based constraints, may confuse the LSTM's learning process.

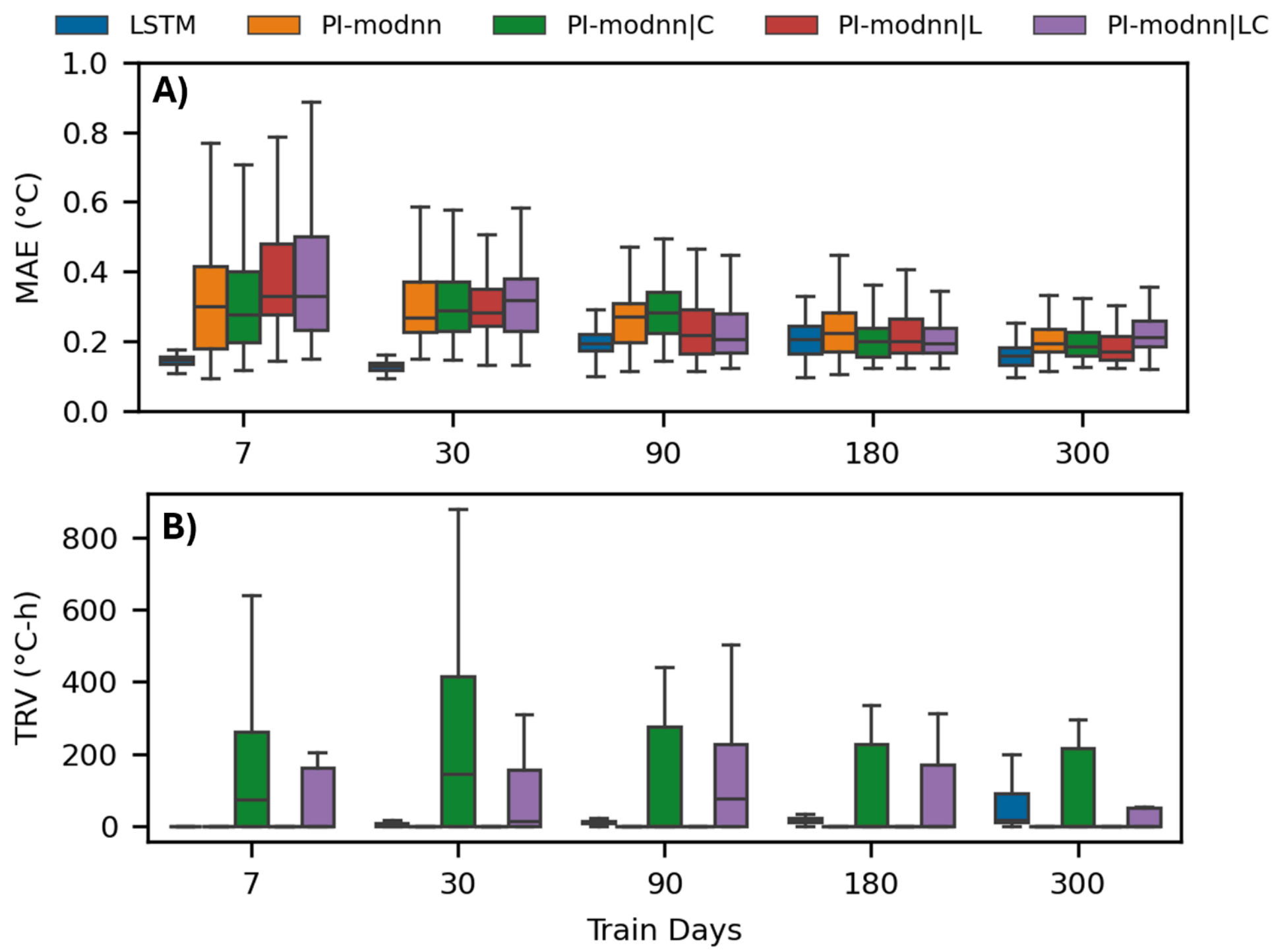

**Fig. 4.** Model performance comparison across different training days. A) Accuracy evaluation. B) Consistency evaluation.

In contrast, PI-ModNN performance improves as the amount of training data increases. This is because the embedded physics rules help the model generalize across seasonal variations. Additionally, epistemic uncertainty also decreased with larger datasets, as shown by a narrower range of prediction errors—indicating the model effectively learns from additional information. However, the aleatory uncertainty shows limited improvement beyond 90 days of training data, likely due to an information gap: although more data were available, the diversity of the dataset did not fundamentally change. For example, the tested space is surrounded by three corridor-facing surfaces, and the absence of adjacent-zone thermal data limits the model's ability to capture unobserved thermal interactions. In other words, adding more redundant data cannot resolve uncertainties arising from unmeasured external influences.

### 3.2. Model physical consistency evaluation

In addition to accuracy, we evaluated the physical consistency of each model as shown in Fig. 4B. For ModNNs with hard constraints (represented in yellow and red), the TRV is consistently zero, indicating strict adherence to physical laws. In contrast, the other models exhibit violations, meaning that they fail to accurately capture the response to HVAC input. And we also notice that the PI-ModNN variants without hard constraints sometimes performed worse than the LSTM baseline in terms of physical consistency. This is likely because the soft constraints (e.g., structural priors and physics-informed losses) used in this study were insufficient to guide training effectively. In some cases, they may have even negatively impacted the optimization landscape, increasing the risk of overfitting and leading to degraded physical behavior compared to the more straightforward LSTM model. This finding highlights the importance of carefully designing and validating physics priors to ensure they provide meaningful guidance during model training.

Interestingly, the response violations of the LSTM model increased with more training data, which may seem counterintuitive. While larger datasets often improve performance, in this case, the broader dataset (300 days) introduced more diverse operating conditions—including both heating and cooling seasons and varied occupancy patterns. The LSTM, lacking physical structure, attempts to learn statistical correlations rather than underlying relationships. As a result, it may overfit to conflicting patterns and become less physically consistent as training data grows more complex.

A detailed example is provided in Fig. 5A and B, where HVAC input ranges from −4 kW (cooling) to +4 kW (heating), shown by a blue-to-red color gradient. The PI-ModNN model responds correctly to HVAC inputs, with temperature decreasing with additional cooling and increasing with additional heating. In contrast, although the LSTM model's predictions align with the measured data, its response offsets significantly from the expected physical behavior. Specifically, the temperature remains unchanged despite variations in HVAC power, indicating that this model is unsuitable for control optimization due to its incorrect response behavior. Further analysis of gradient distributions (*Figure A5*) shows that PI-ModNN maintains non-negative gradients with respect to both control and disturbance inputs, reflecting its physics-consistent structure. The LSTM model lacks such guarantees, often violating expected physical behavior.

### 3.3. Value of prior knowledge evaluation

To evaluate the contribution of physics priors to model performance, we assess their impact on both accuracy and physical consistency, as shown in Fig. 6A. Compared to LSTM, none of the physics-informed rules provide a clear advantage. Under limited training data (e.g., 7 days), the hard constraint slightly improved model performance, whereas the structural modification led to a performance decline—likely because the modified architecture requires more data to learn effectively. As the training data increases, the median value of knowledge is close to zero, indicating that under normal testing conditions, most model constraints offer limited benefits. This is due to the restricted solution space can lead to a performance drop as discussed in previous studies [33,41].

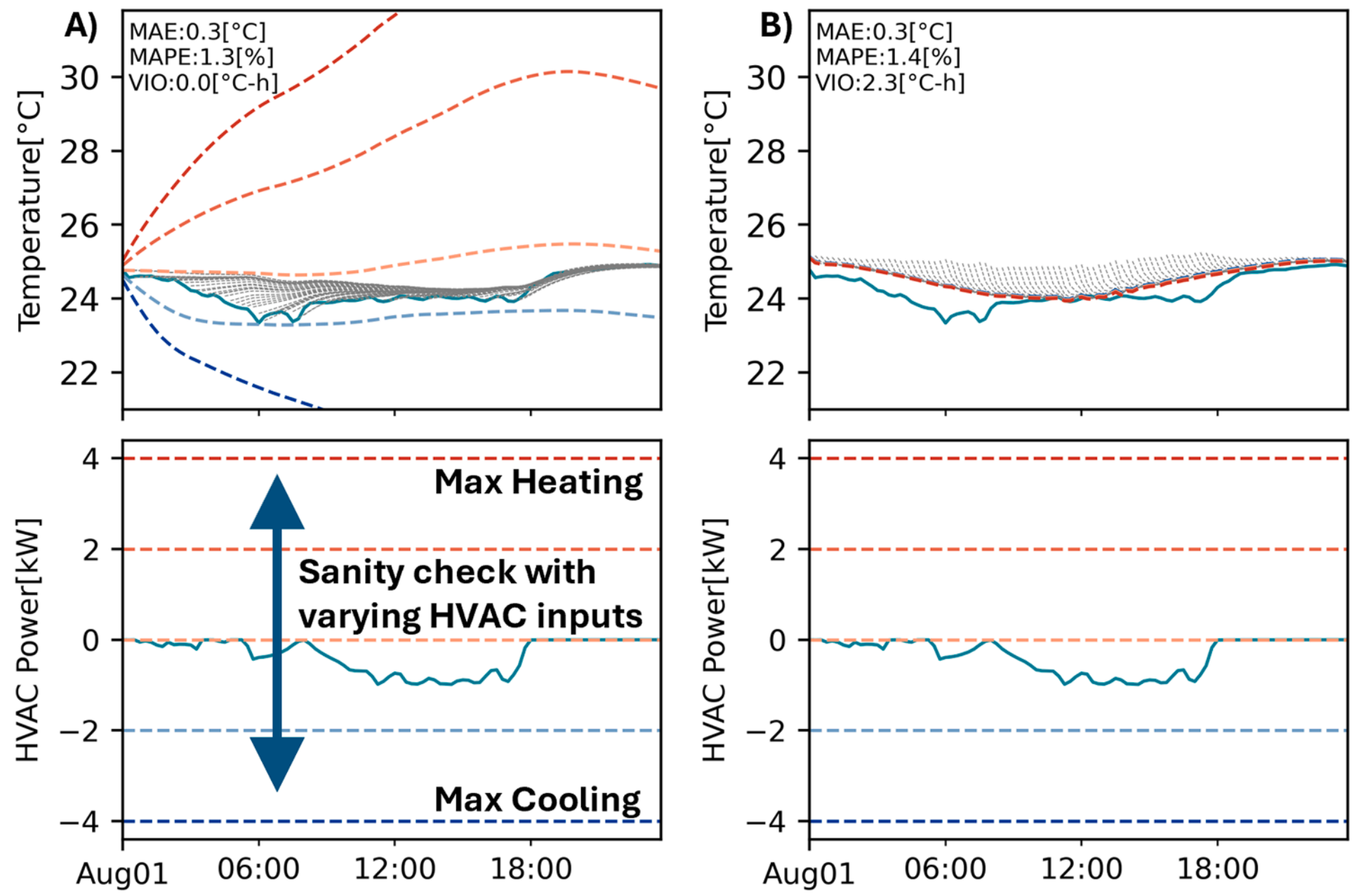

**Fig. 5.** Example of temperature response under varying levels of HVAC input. A) PI-ModNN model. B) LSTM model.

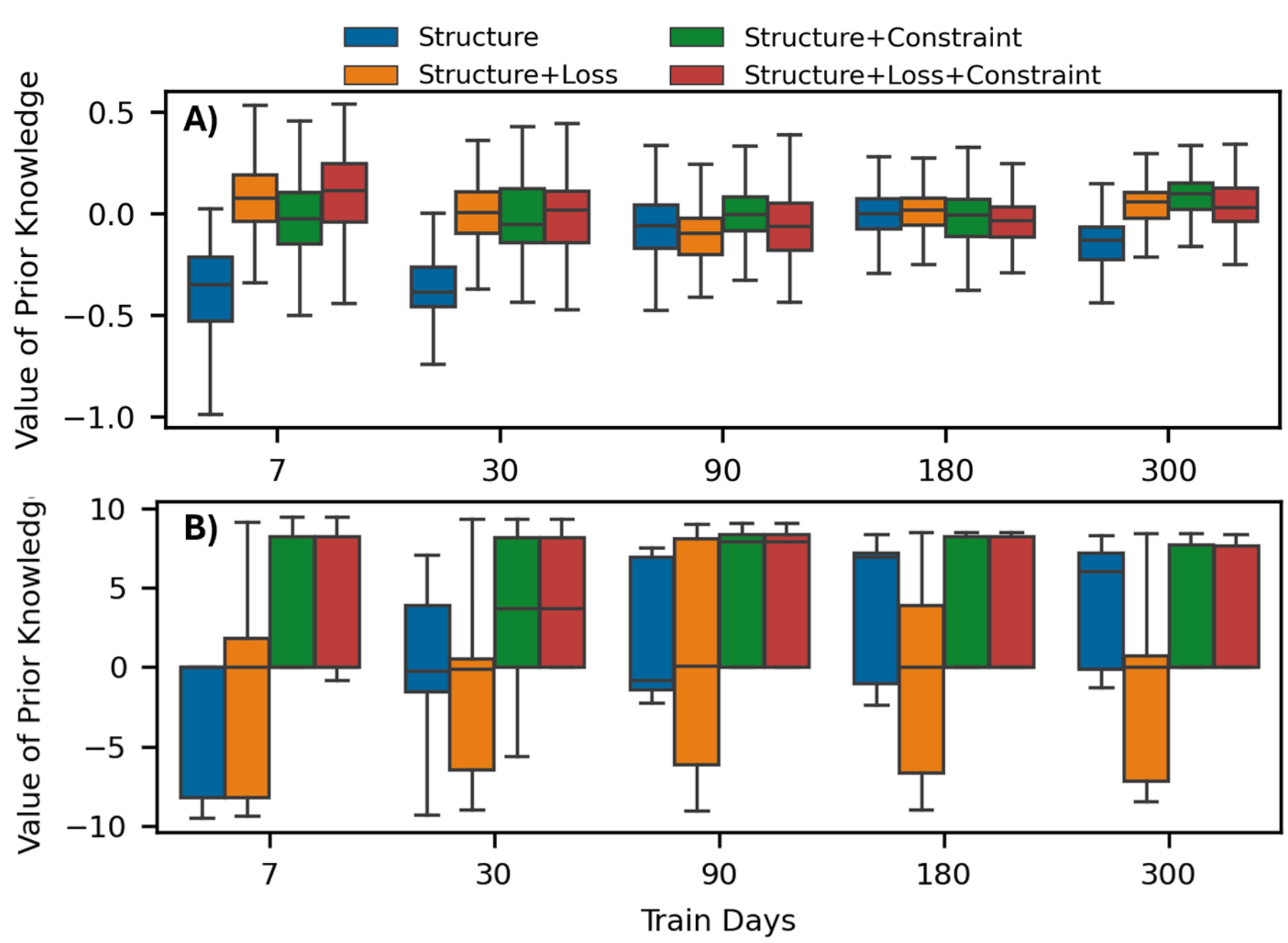

**Fig. 6.** Evaluation of the impact of prior knowledge. A) Effect of prior knowledge on model accuracy. B) Effect of prior knowledge on model consistency.

In contrast, physics priors had a more significant effect on physical consistency (Fig. 6B). Structural modifications initially performed poorly with small datasets (e.g., 7 days) but yielded consistent improvements as more data became available. This suggests that physics-informed architectures, which guide the model to learn temperature changes rather than static mappings might require sufficient data to be effective. Once trained, they improve consistency by aligning learning with underlying physical relationships.

For the physics-informed loss function, we observe negative impacts in most cases. This finding suggests that adding physical priors does not always lead to improved performance. To better understand the potential reasons, we categorize physics priors into two types:

- **Knowledge-based priors:** Derived from governing equations, these offer strong physical validity but may degrade performance due to complex loss balancing, soft constraint violations and sensitivity to boundary/initial conditions, especially under real-world uncertainty or non-homogeneous conditions.
- **Empirical-based priors:** Based on empirical experience (e.g., fluctuation-based loss used in this study). However, this knowledge

could be biased and does not always guarantee improved model performance due to our limited understanding of how learning unfolds in high-dimensional latent spaces.

Models with hard physical constraints (green and red in Fig. 6B) demonstrate significant improvements in model consistency. This is because hard constraints explicitly enforce physical laws, restricting the solution space to physically feasible regions and preventing unrealistic outputs. However, depending on how they are formulated, hard constraints may introduce sharp boundaries in the optimization landscape, potentially complicating training.

In contrast, soft constraints encode physics as regularization terms in the loss function. They offer greater flexibility and has a smoother optimization landscape compared to hard constraint, potentially accelerating convergence. However, they do not guarantee constraint satisfaction and require careful tuning of weighting factors to balance accuracy and consistency. Their effectiveness is often uncertain and task-dependent.

In this study, the hard constraint improved model consistency without compromising training, highlighting the importance of selecting and validating appropriate priors. Not all priors are beneficial, and their value depends on alignment with the learning objective and the available data.

### 3.4. Experiment performance

We first compare the control performance of the DRL agent using PI-ModNN and LSTM as environment models for a representative summer day, as shown in *Figure A6*. Under the PI-ModNN environment, the space air temperature is well maintained near the upper bound of the setpoint range, indicating successful policy learning. This aligns with simulation results where the agent achieved a score of 5. In contrast, the LSTM-based environment results in poor control performance, with the DRL agent scoring only -50. Notably, the LSTM-based controller mistakenly applies heating while the temperature continues to decrease, highlighting a misrepresentation of thermal dynamics. The root cause of this issue is the LSTM fails to capture the correct thermal response—for example, predicting a temperature increase under maximum cooling (*Figure A6*). This discrepancy leads to incorrect feedback for the DRL agent, ultimately preventing it from learning an effective control policy.

We then deployed the DRL agent trained with the PI-ModNN environment. Experimental results (Fig. 7) show that compared to the baseline, ModNN-DRL:

- Reduces temperature violation from **1.04 °C-h** to **0.07 °C-h**,
- Achieves a **31.4%** reduction in coil-side HVAC load by leveraging free cooling,
- Shifts **28.4%** of peak load away from morning occupancy hours due to smooth reward shaping

However, we observe a clear performance gap when comparing it to the simulation results, as shown in Fig. 8. In the simulation, the coil load remains close to zero by fully utilizing free cooling from the outside air, thereby saving energy while maintaining good thermal comfort. In contrast, the experiment consistently shows coil load consumption even when the space temperature does not reach the upper bound of the setpoint.

To investigate, we check the detailed daily HVAC operation (Fig. 9A, February 11$^{th}$). Cooling continues even the space air temperature has not reached the upper bound of the setpoint. This behavior explains the mismatch between the simulation and experimental results, limiting the potential for energy savings. This is because ModNN was trained on data collected before January 15, 2025, when internal heat gain was generated by real occupancy behavior. But during the DRL implementation stage, since all students moved out and the space was unoccupied, heat lamps were used to replace internal heat gains. This changed internal thermal dynamics is not reflected in the training dataset, and the amount of heat generation is under-represented, causing ModNN to overestimate space air temperature, as shown in *Figure A7*.

To address this mismatch, the ModNN model was retrained using updated data collected after January 15, when internal heat gains were provided by heat lamps. Based on this new environment model, the DRL agent was also retrained using data from the same period. The updated performance is shown in Fig. 9B. On a representative day, March 11, the space air temperature exceeded the setpoint between 11:00 and 18:00, lasting for 5 hours with a peak violation of 1.6 °C. During this period, the HVAC system was expected to provide cooling. However, the DRL agent incorrectly shut off cooling, resulting in thermal discomfort. This control failure is likely due to the limited training dataset—approximately one month of data between January and February—during which ambient temperatures were lower. The significantly higher outdoor temperature on March 11 falls outside the agent's training distribution, leading to poor generalization under unseen conditions.

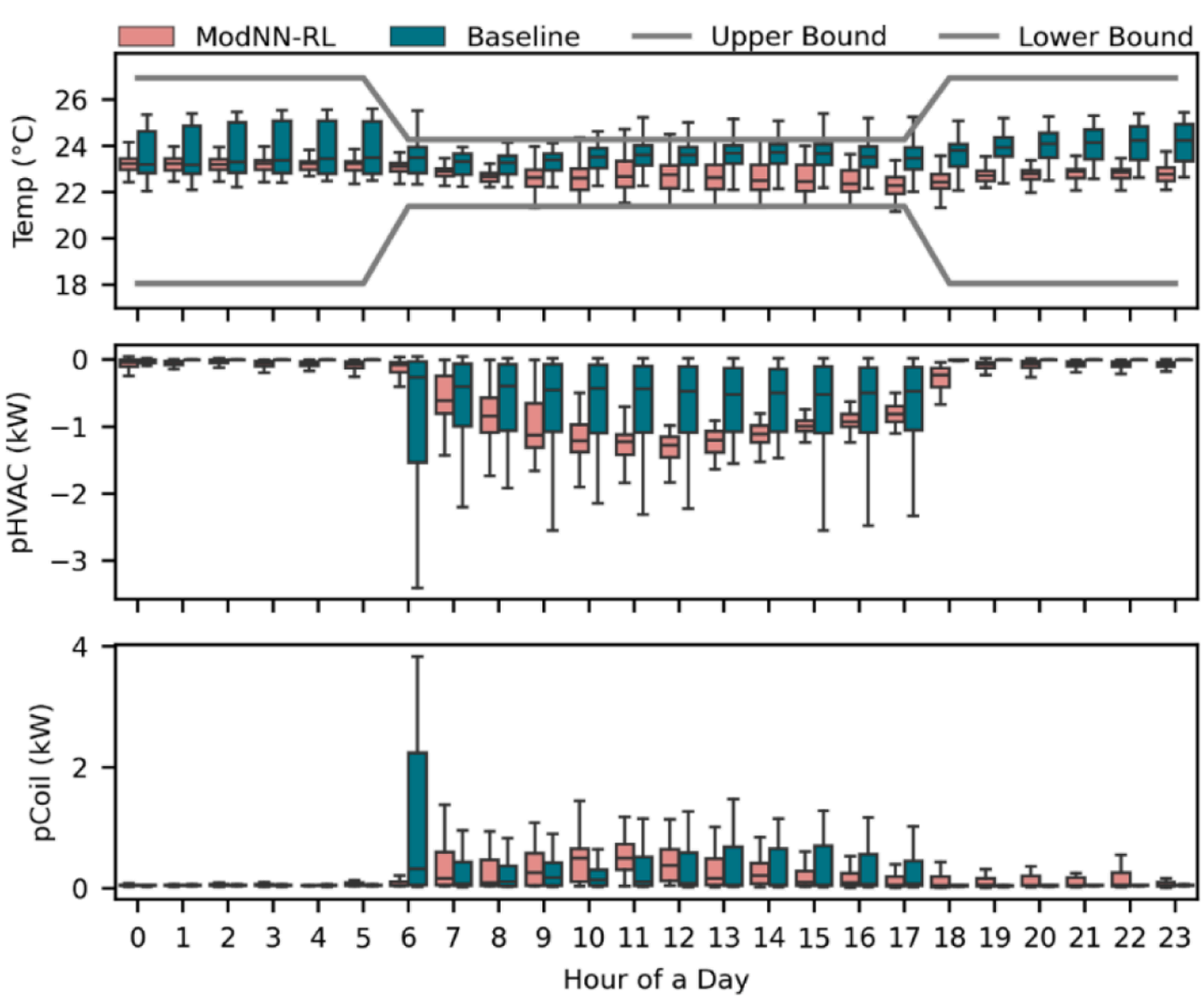

**Fig. 7.** Thermal comfort and HVAC energy comparison of ModNN-RL and Baseline, experiment result.

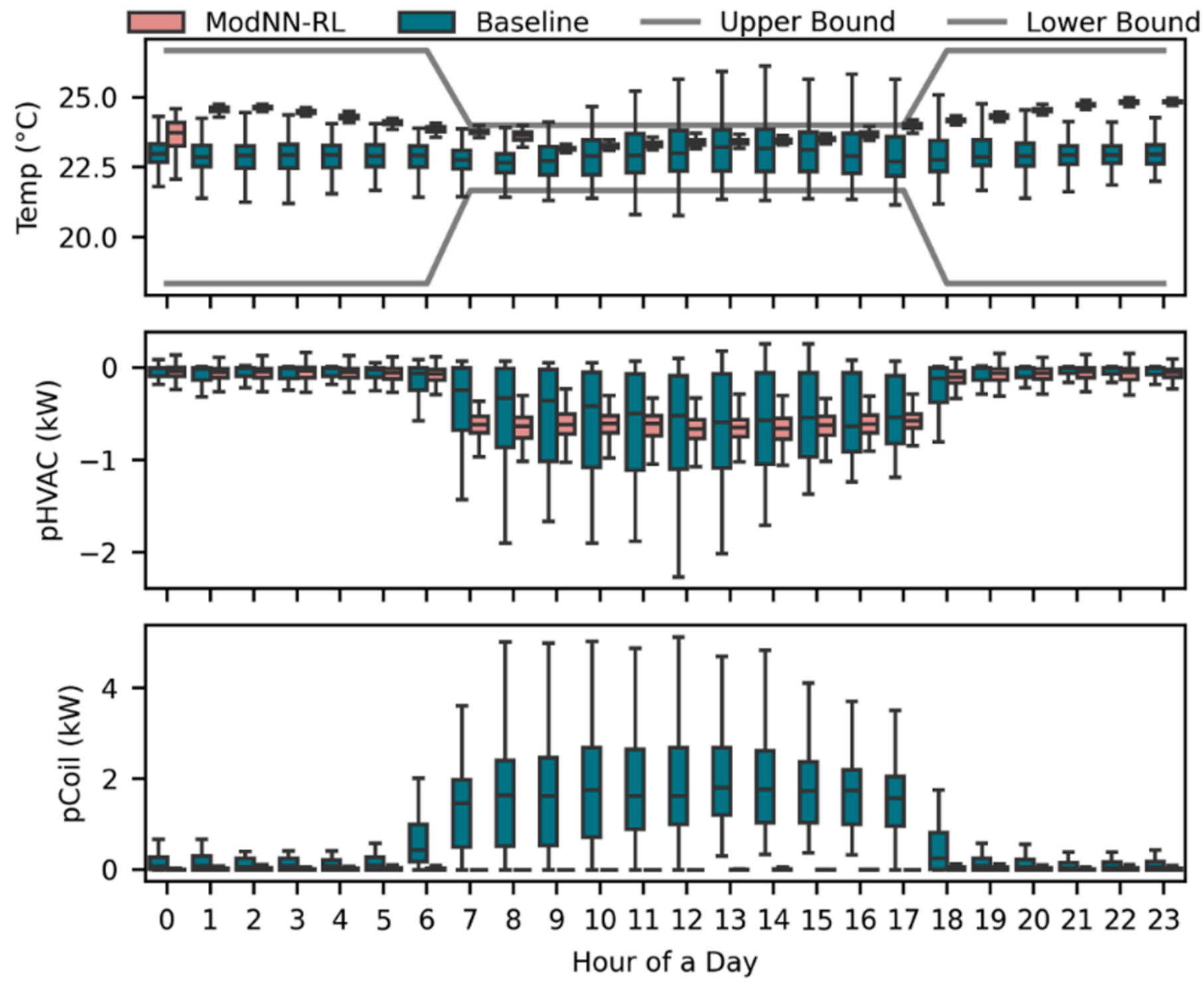

**Fig. 8.** Thermal comfort and HVAC energy comparison of ModNN-RL and Baseline, simulation result.

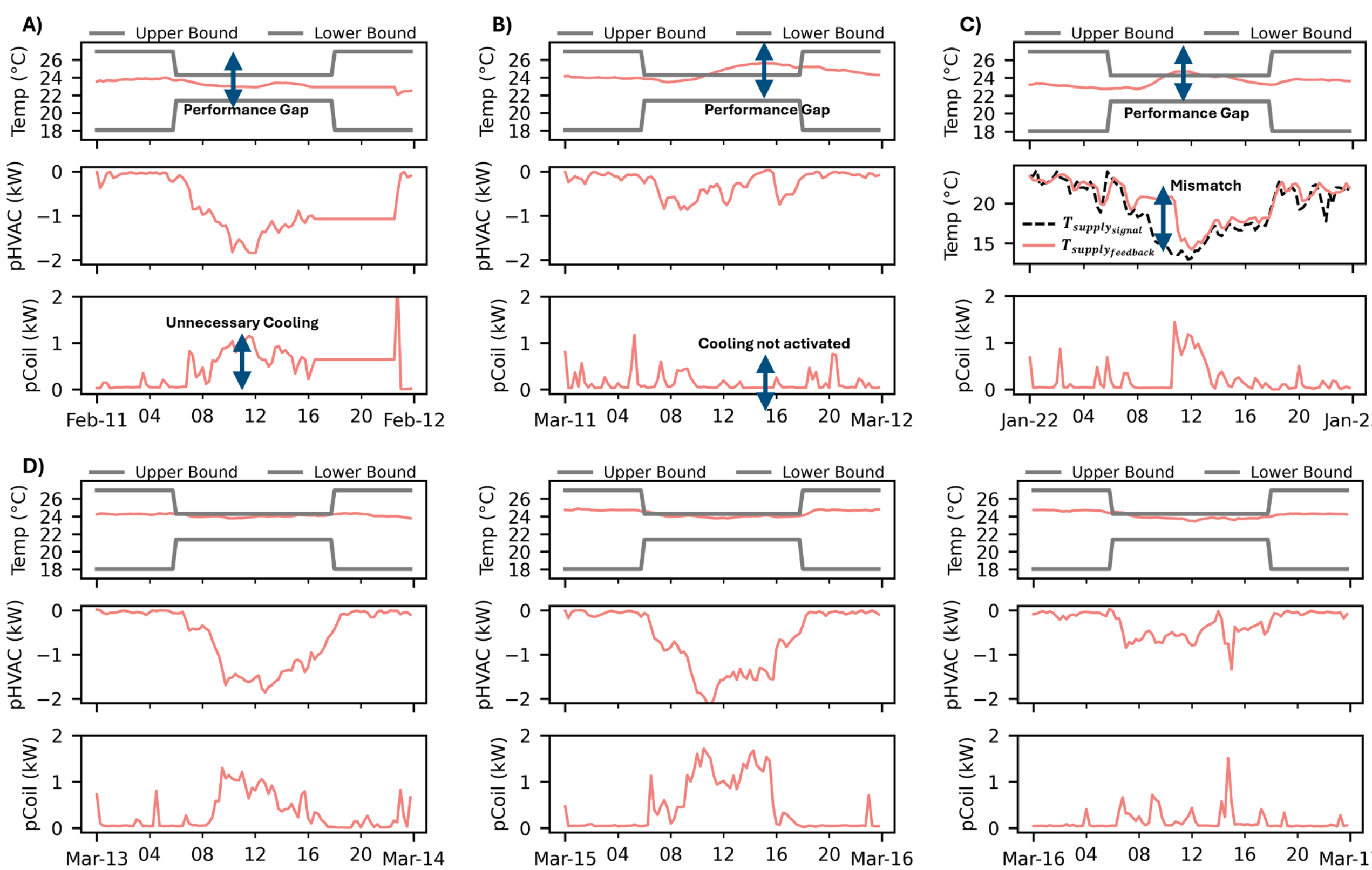

**Fig. 9.** Representative experimental results from the DRL experiments on selected days, showing space air temperature, HVAC thermal load, and coil-side load.
A) Both DRL and PI-ModNN were trained using data with the old internal dynamics.
B) Both DRL and PI-ModNN were trained using data with the new internal dynamics.
C) A short period of temperature violation occurred due to a mismatch between the control action and system response.
D) PI-ModNN was trained on data with the new internal dynamics, while DRL was trained on all historical data.

To fix this problem and improve the generalization of DRL agent, we retained the DRL agent using all historical data (August 2023 to March 2025) using same PI-ModNN environment (trained on data after January 15, 2025). The updated results, shown in Fig. 9D, demonstrate that the space air temperature is now closely aligned with the setpoint upper bound, effectively reducing unnecessary HVAC energy consumption. Additionally, the temperature violation reduced by 4.3°C-h compared to Fig. 9B.

Another observation is that electrical energy savings do not directly correspond to HVAC thermal load reduction. As shown in Fig. 7 and Fig. 8, although the DRL results in a slightly higher thermal load, the actual coil load is reduced—likely due to the use of free cooling. This highlights a key limitation in many simulation-based studies, where thermal load is often assumed to be directly controllable. In real-world systems, however, the actual control variables are typically the supply air flow rate or the supply air temperature setpoint. Neglecting the dynamics and operational constraints of the HVAC system can lead to suboptimal or even infeasible control strategies. For example, as shown in Fig. 9C, the HVAC system was expected to provide additional cooling by decreasing the supply air temperature. However, due to a possible capacity limitation, the actual supply air temperature was higher than the control signal provided by the DRL agent. This leads to a short temperature violation of the upper bound of the setpoint. Such issues could be mitigated by integrating a more detailed HVAC system model, which is essential for developing advanced controllers that are both effective and practically deployable.

## 4. Discussion

### *4.1. A general guideline for using data-driven models in building control*

This study proposes a four-step evaluation framework to support the integration of data-driven models into advanced building control

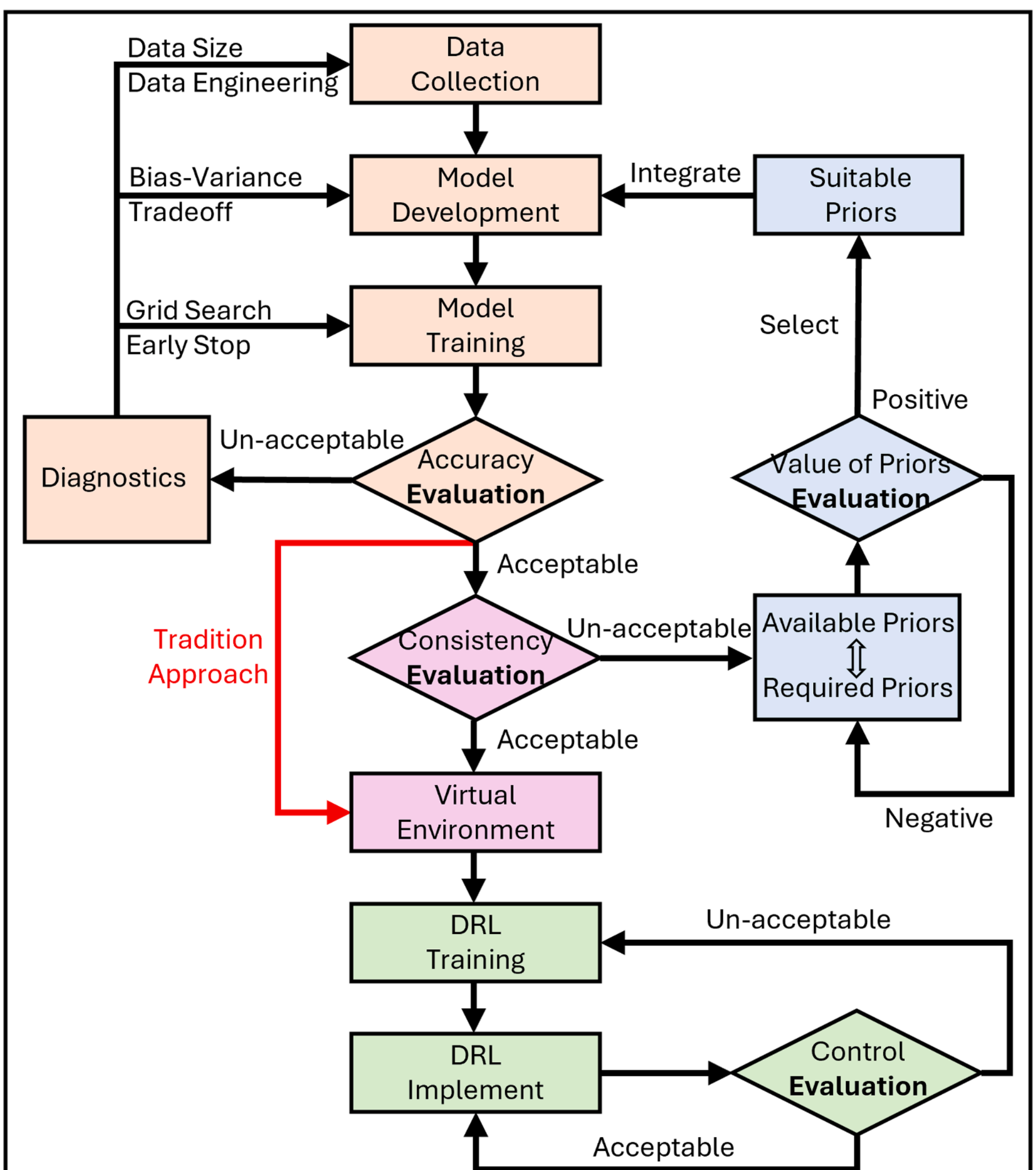

**Fig. 10.** Most current data-driven building thermal models are evaluated solely based on prediction accuracy metrics (e.g., MAE, RMSE) while physical consistency is often overlooked. However, models that predict well but respond incorrectly to control inputs can lead to serious control failures.

systems, as shown in Fig. 10.

Particular for control-oriented applications, models must be both accurate and physically responsive. To address this, we introduce a consistency evaluation framework, which quantifies model behavior using a response violation metric. This enables users to assess whether the model produces physically plausible responses to varying control inputs.

Nonetheless, we emphasize that evaluating the control performance and physics consistency of surrogate models remains an open research challenge. In data-driven modeling, performance can vary a lot due to differences in hyperparameters, training epochs, and initializations. These sources of uncertainty can significantly affect the model's response behavior. Even models with similar MAE may show different control responses and physics consistency. Further research should explore potential evaluation solutions, which can be integrated into our proposed four-step framework to support more robust and reliable model selection.

Then another question is: "Does this mean purely data-driven models cannot be used for building control optimization?" or "Do all data-driven models require the incorporation of physical priors?". The answer is no. If a purely data-driven model passes the consistency evaluation, it can be used directly for control optimization without additional physical priors. For example, in our experiments, some LSTM models successfully captured HVAC–temperature relationships and exhibited zero response violations. However, due to the lack of hard constraints, their performance was unstable—working well with certain training parameters during some training epochs but failing in others. As shown in *Figure A8*, we present two examples where, after multiple trials, an LSTM model achieved zero temperature response violations, also indicating this matrix can be further refined. However, the reliability of such models relies on obtaining a "lucky" set of training parameters, which is difficult to consistently achieve. Another example is showed in *Figure A9*, where although the accuracy loss converges, the consistency loss of the LSTM model fluctuates, indicating that its response remains physically inconsistent. These findings suggest that the consistency of a purely data-driven model is not guaranteed and, in most cases, cannot be satisfied. Therefore, appropriate physical priors need to be carefully incorporated to address this issue.

For models that fail the consistency test, physical priors can be incorporated in various forms (e.g., structure, constraints, loss terms). However, user-defined priors can introduce bias due to incomplete understanding or limited observations. Therefore, we propose a value-of-priors evaluation framework to help users identify which priors are most beneficial. The selection of priors should depend on both domain knowledge and the required level of physical consistency for a given task, balancing the trade-off between model flexibility and reliability.

Finally, even a physically consistent model must undergo an additional verification step before DRL deployment. Since DRL itself is data-driven, its generalization capability must be tested thoroughly. Key open questions include:

- How can we more systematically evaluate the physical consistency of a data-driven model?
- Can we incorporate physical priors to DRL agents and improve their generalization ability?
- How can we ensure the safety of the learned policy?
- When should the DRL agent be updated using newly available real-world data?

These questions point to promising directions for future work in developing robust and reliable data-driven control frameworks for smart buildings.

### 4.2. Module selection and development

A modularized model offers great flexibility, enabling users to customize each module for specific objectives. Here, we briefly discuss key considerations for module selection and development, along with open questions that remain to be addressed:

#### 4.2.1. External heat transfer module

An RNN-type structure is used to model external heat transfer, with hidden states designed to capture the thermal inertia of the building envelope. The structure selection depends on the modeling objective: while LSTM/GRU address gradient vanishing issues, vanilla RNNs are more suitable for enforcing positive gradient constraint[33], despite this constraint may not be necessary sometime. For example, ambient temperature typically has a narrow distribution, and a purely data-driven model may be sufficiently accurate. Alternative approaches have been explored. For instance, Drgoňa et al. [45] developed a fully connected neural network without embedding physics-based constraints. Di Natale et al. [40] adopted a linear model, enabling the control problem to be framed as a convex linear program which can be solved with high efficiency. However, this simplification neglects envelope thermal inertia, potentially introducing bias for heavy-mass buildings. In comparison, this study couples state variables with the disturbance model, which improves accuracy but increases the model complexity. There remain open questions on how to more accurately model external heat transfer, such as the integration with lumped RC model. We also notice that incorporating time-related features can improve prediction accuracy, however, it may compromise physical consistency. Indoor temperature is strongly influenced by scheduled HVAC control, and time-based features may lead to overfitting rather than learning the true heat transfer process. Therefore, careful selection of model inputs is crucial to ensure both predictive accuracy and physical interpretability.

#### 4.2.2. Internal heat transfer module

We use an MLP generates schedule-like outputs (0–1 via sigmoid) from time features, which are combined with normalized occupancy and passed through a scaling layer to produce internal gains (occupants, lighting, equipment). This configuration only captures convective gains, where the lag effects of radiation require the use of historical data through lookback windows to be accurately represented.

#### 4.2.3. HVAC module

We use an MLP maps control inputs to HVAC heat transfer for air-based systems. For systems with thermal lag (hydronic or radiation-based systems), RNNs or MLPs with lookback windows will be better. When HVAC load data is missing (e.g. dataset without air flow reading), alternative inputs like supply air temperature or damper positions can be used to infer HVAC impact implicitly using a separately HVAC module, improving scalability in sensor-limited buildings.

#### 4.2.4. Residual module

This module estimates unknown thermal capacity (assuming well-mixed air and neglecting surface storage) using an MLP. For greater accuracy, it can be upgraded to RNNs, or MLPs with a lookback window, or higher-order integration methods like Runge-Kutta [46] or neural ODEs [47].

#### 4.2.5. Open questions

Several open questions remain in the development and integration of modular components. 1) how can each module be trained effectively in partially observable systems where the label is limited? 2) determining the appropriate degree of physics integration remains challenging, as it depends on the specific modeling objective, the availability of data and domain knowledge, and the trade-off between interpretability and learning capacity. 3) efficient module integration is non-trivial—outputs from different modules may vary in scale or format, complicating direct combination. One strategy is to concatenate outputs and process them through an additional data-driven layer, which enhances flexibility but reduces interpretability; alternatively, aggregating outputs (e.g., summing latent heat flows) requires careful normalization to maintain physical consistency. 4) it remains unclear how to define a unified model structure that supports pre-training of individual modules and their seamless composition in integrated systems.

### 4.3. Generalization of DRL agent

As shown in Fig. 9, the DRL agent did not perform as expected when trained with a very limited dataset. Then the question is: how can the generalization ability of DRL agents, which are also data-driven models, be improved in the future? We discuss three potential solutions below:

- **Data augmentation**
  Once the surrogate environment is trained, synthetic data can be applied to enrich the training dataset of DRL. For example, occupancy patterns can be altered by substituting unoccupied periods with occupied profiles or introducing some random noises to the weather data to create diverse scenarios. It can expose the DRL agent to diverse state-action combinations and improves its robustness and generalization to handle variability in real-world deployment.
- **Hybrid learning**
  In addition to training through interactions with the simulated environment, DRL agents can be pre-trained on historical building data to learn fundamental control behaviors. This pre-training phase provides a robust initial policy before deployment, reducing poor early decisions. In addition, the agent can continue learning and refining its control policy online through further interaction with either the simulation or the physical system.
- **Physics-informed DRL design and model based DRL**
  Incorporating physics constraints could improve DRL consistency and data efficiency. For example, feasible operating bounds can be enforced via hard constraints on the policy network. With a differentiable PI-ModNN, model-based DRL becomes feasible—allowing agents to simulate future states and improve learning through gradient-based updates. This reduces data needs, improves generalization, and supports safe deployment in real buildings.

### 4.4. Experimental variability due to occupancy behavior

Occupant behavior significantly influences building thermal dynamics, particularly in small office. As discussed in Section 4.3, after occupants moved out and heat lamps were used to simulate internal heat gains, the original model trained on the earlier occupancy data consistently overestimated the space temperature, leading the DRL agent to apply unnecessary cooling. This is because the heat lamps provided less heat than actual occupants and plug loads, resulting in a mismatch between the model's predictions and the real thermal response. Once the model was retrained using data collected after the substitution with heat lamps, the performance gap was resolved. This highlights the importance of ensuring that training data maintains consistent characteristics to avoid misleading the learning process.

### 4.5. Limitations and future study

In this study, we quantified physical consistency both qualitatively and quantitatively. However, this metric represents only the minimal requirement for applying a data-driven model for control purposes. In other words, the current method focuses on the sign of the control response rather than the exact magnitude. An open question remains: how can we more accurately evaluate physical consistency? For example, as shown in *Figure S12*, although the consistency violation is zero, the system response appears unrealistic. To better quantify consistency, it might be necessary to collect real-world sanity check data—such as the change in space temperature under different levels of HVAC input—or use simulation data for verification.

Another limitation is that the marginal contribution of each physical prior was not evaluated independently. In this study, we tested the combined effect of multiple priors. Future work could apply a more systematic sensitivity analysis to assess the individual contribution of each physical prior.

Additionally, in this study, physics priors were incorporated at the beginning of training. Future work could explore the impact of introducing physics priors dynamically—e.g., adding them midway through training or gradually adjusting their weight—to better understand how the timing and strength of prior integration influence model performance and generalization.

Another potential direction is the integration with neural ordinary differential equations[35,36], which provide a continuous-time modeling framework that is naturally aligned with physical dynamics. By embedding physics priors into the neural ODE structure, it may be possible to improve both the interpretability and temporal consistency of the learned model.

## 5. Conclusion

In this study, we developed a Physics-Informed Modularized Neural Network that integrates physics-informed model structures, loss functions, and constraints for building dynamic modeling. The model's prediction accuracy was evaluated over one month with varying training data sizes, achieving an averaged MAE ranging from 0.2 to 0.33°C. Additionally, we evaluate its physical consistency using a new defined evaluation matrix based on temperature response violations. Furthermore, we proposed an evaluation framework to assess the contribution of individual physics priors based on rule importance.

The incorporated physical priors lead to slight model performance drop in terms of model accuracy due to the limited solution space. However, the physics-informed model structure enhanced model consistency when the training data exceeded 30 days. While adjusted loss functions negatively impacted model consistency, hard constraints significantly improved it, indicating that appropriate selection of physical priors is essential for Physics-Informed Machine Learning development.

We then integrated PI-ModNN as a virtual environment to train the deep reinforcement learning agent, which was implemented in a small office building for three months. The deep reinforcement learning agent demonstrated an energy savings potential of over 30%. Finally, we provide a general guideline for integrating data-driven models with advanced building control using a four-step evaluation framework. This study contributes to the reliable implementation of data-driven advanced building control and offers valuable insights for future researchers in this field.

## CRediT authorship contribution statement

**Zixin Jiang:** Writing – review & editing, Writing – original draft, Visualization, Validation, Software, Methodology, Investigation, Formal analysis, Conceptualization. **Xuezheng Wang:** Writing – review & editing, Writing – original draft, Software, Methodology, Conceptualization. **Bing Dong:** Writing – review & editing, Writing – original draft, Supervision, Resources, Project administration, Methodology, Funding acquisition, Conceptualization.

## Declaration of competing interest

The authors declare that they have no known competing financial interests or personal relationships that could have appeared to influence the work reported in this paper.

The author is an Editorial Board Member/Editor-in-Chief/Associate Editor/Guest Editor for this journal and was not involved in the editorial review or the decision to publish this article.

## Supplementary materials

Supplementary material associated with this article can be found, in the online version, at doi:10.1016/j.adapen.2025.100237.

## Appendix

1. Sensitivity Analysis of Different Loss Mixing Ratios
Fig. A1, A2, A3, A4, A5, A6, A7, A8, A9, Table A1

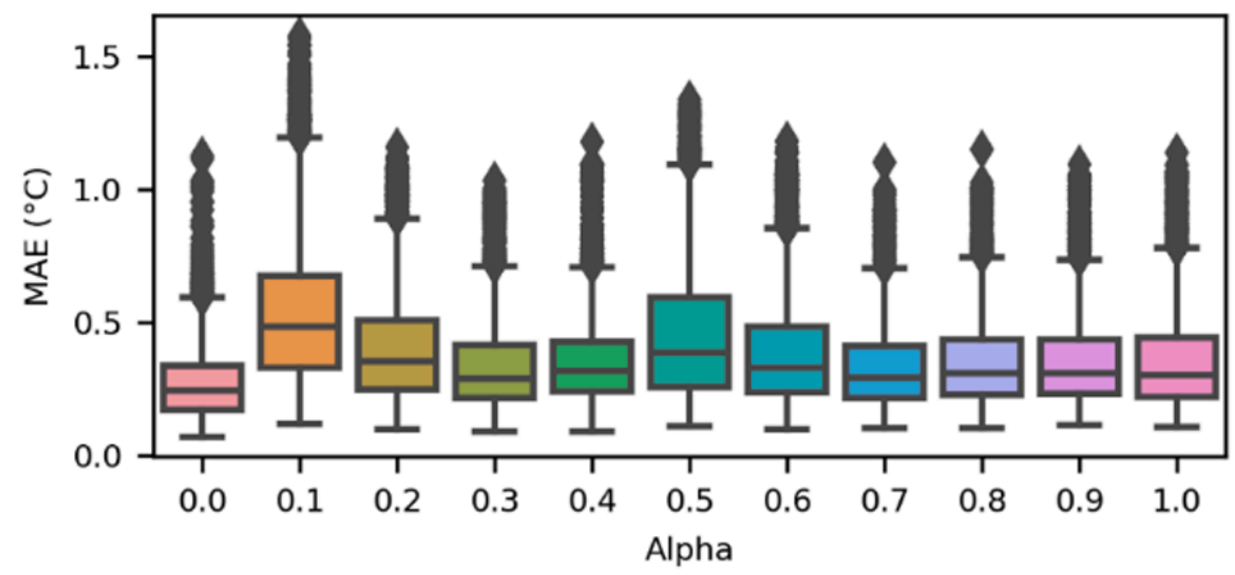

**Fig. A1.** Model accuracy under different ratios of incorporated physics-informed loss, indicating that this specific loss did not lead to improvement in this case.

2. Impact of Teacher-force Strategy on Model Training

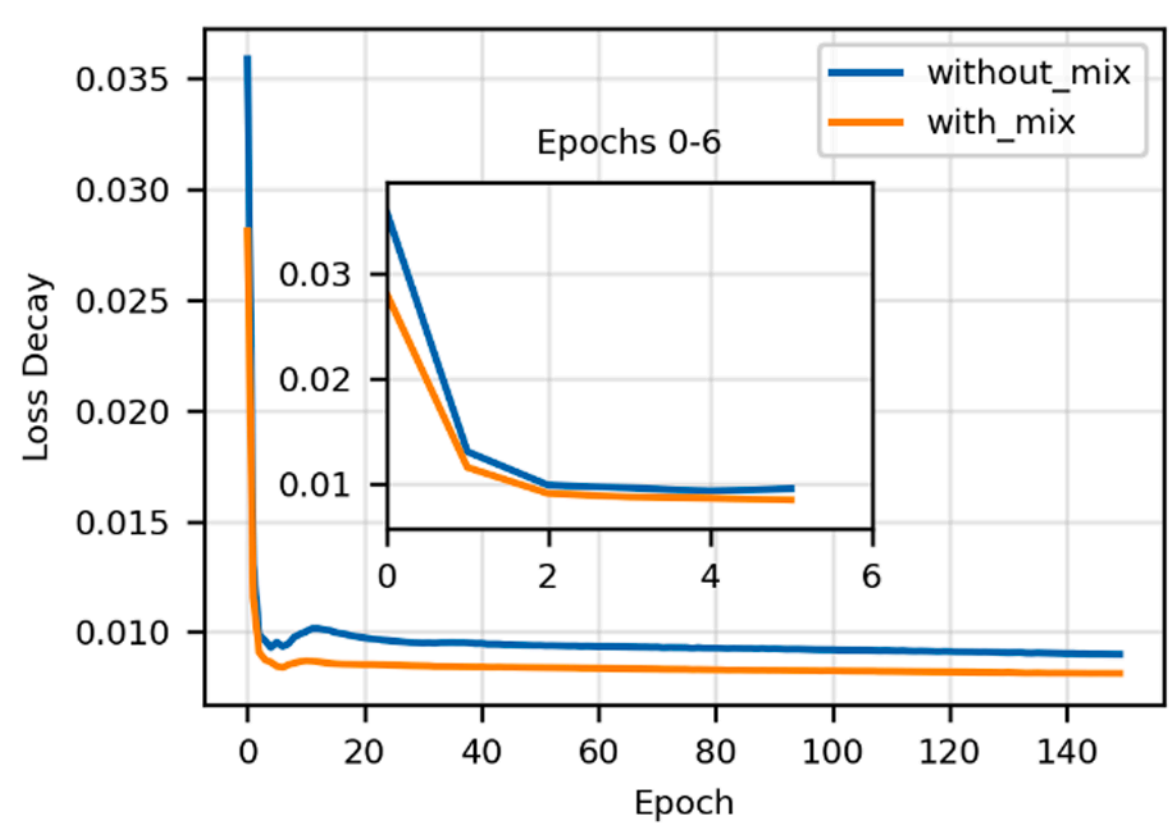

**Fig. A2.** Loss decay with and without teacher-force strategy. Mixed data slightly improves performance, since the injection of ground truth data at encoder stage helps model focus on learning the dynamic model.

3. Impact of Training Strategy and Environment Model on DRL Learning Performance

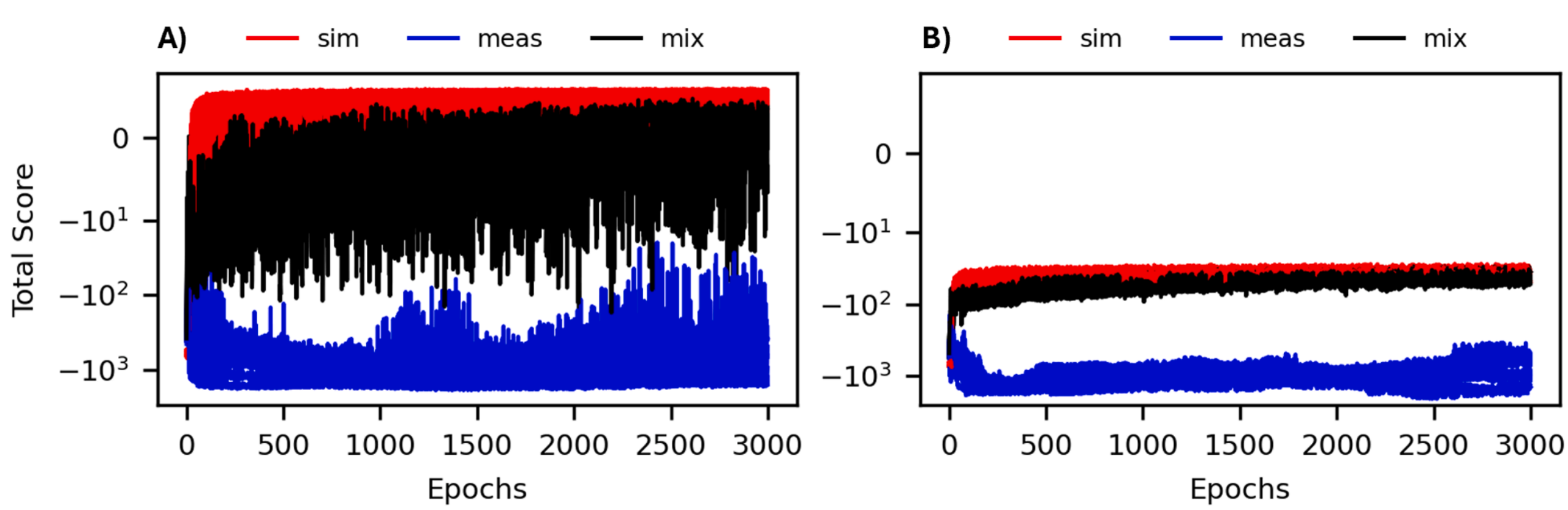

**Fig. A3.** Learning curves of the DRL agent under different training configurations. A) PI-ModNN environment, B) LSTM environment.

The red line represents training through interaction with the simulation environment; the blue line denotes training using only historical measured data; and the black line indicates training with a mix of both. Each curve is averaged over five training runs to improve representativeness. Training on measured data only limits the agent to follow baseline control behavior in the dataset, restricting exploration and resulting in poor learning performance (converging to a low reward of approximately –1000). In contrast, training with the interactive PI-ModNN environment enables effective policy exploration and learning, achieving a much higher reward (around 5). The mixed-data approach yields intermediate performance, primarily due to (1) distributional differences between simulated and measured data, and (2) inconsistencies in control policies across data sources. Moreover, the PI-ModNN environment outperforms the LSTM-based one significantly. While the LSTM model shows good predictive accuracy on standard metrics, it fails to capture accurate system responses to control inputs. As a result, the DRL agent trained in the LSTM environment converges to a suboptimal policy (reward ≈ –50), indicating ineffective learning.

4. Model Performance Evaluation Over a Two-Week Period in August

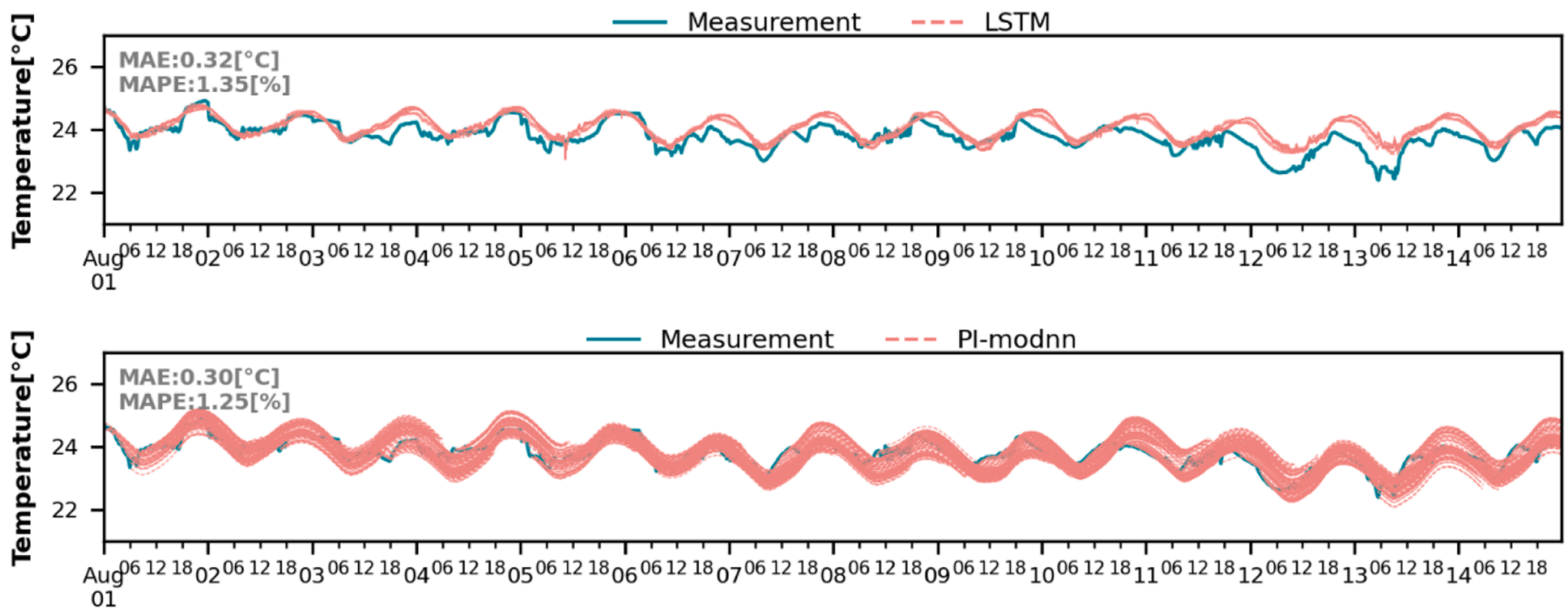

**Fig. A4.** . One-day-ahead temperature predictions by the LSTM model (top) and PI-ModNN (bottom) for the first two weeks of August. The green line represents the measured space air temperature over a 24-hour period, while the red dashed line indicates the predicted temperature. Predictions are plotted at 15-minute intervals.

**Table A1**
. Model prediction error in mean absolute error C.

| | 7 Days | 30 Days | 90 Days | 180 Days | 300 Days |
|---|---|---|---|---|---|
| LSTM | 0.14 | 0.13 | 0.19 | 0.21 | 0.16 |
| PI-ModNN | 0.3 | 0.27 | 0.27 | 0.22 | 0.19 |
| PI-ModNN\|C | 0.28 | 0.29 | 0.28 | 0.20 | 0.18 |
| PI-ModNN\|L | 0.33 | 0.28 | 0.22 | 0.21 | 0.17 |
| PI-ModNN\|LC | 0.33 | 0.32 | 0.21 | 0.19 | 0.21 |

5. Model Gradient Evaluation

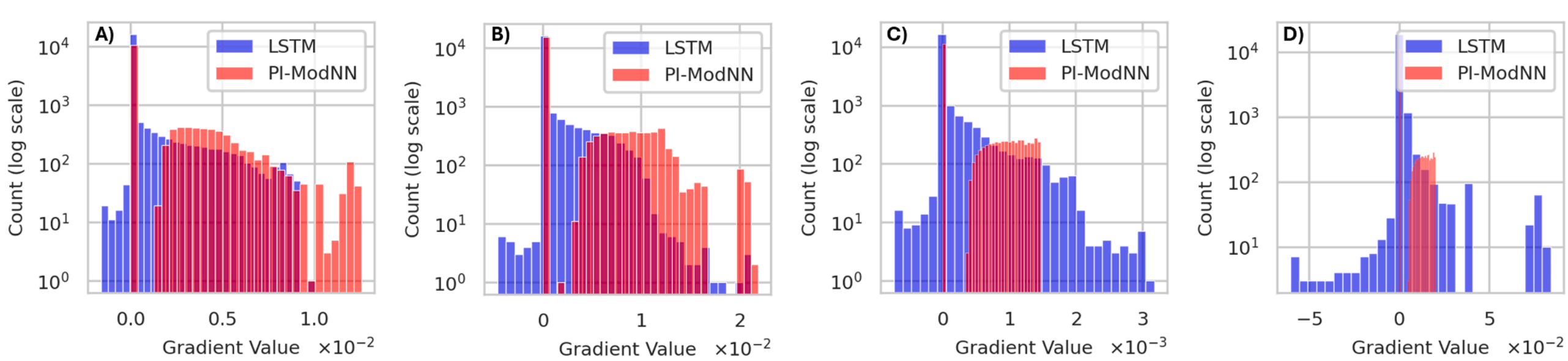

**Fig. A5.** Comparison of gradient distributions between PI-ModNN and LSTM with respect to different input variables. The orange histogram represents PI-ModNN, and the blue histogram represents LSTM. PI-ModNN maintains non-negative gradients with respect to both control and disturbance variables, reflecting its physically consistent structure. In contrast, the LSTM model does not ensure this property, which may result in non-physical responses. A) Solar radiation. B) Ambient temperature. C) Occupancy. D) HVAC.

6. One Day Example of Control Performance Using PI-ModNN and LSTM as Environments

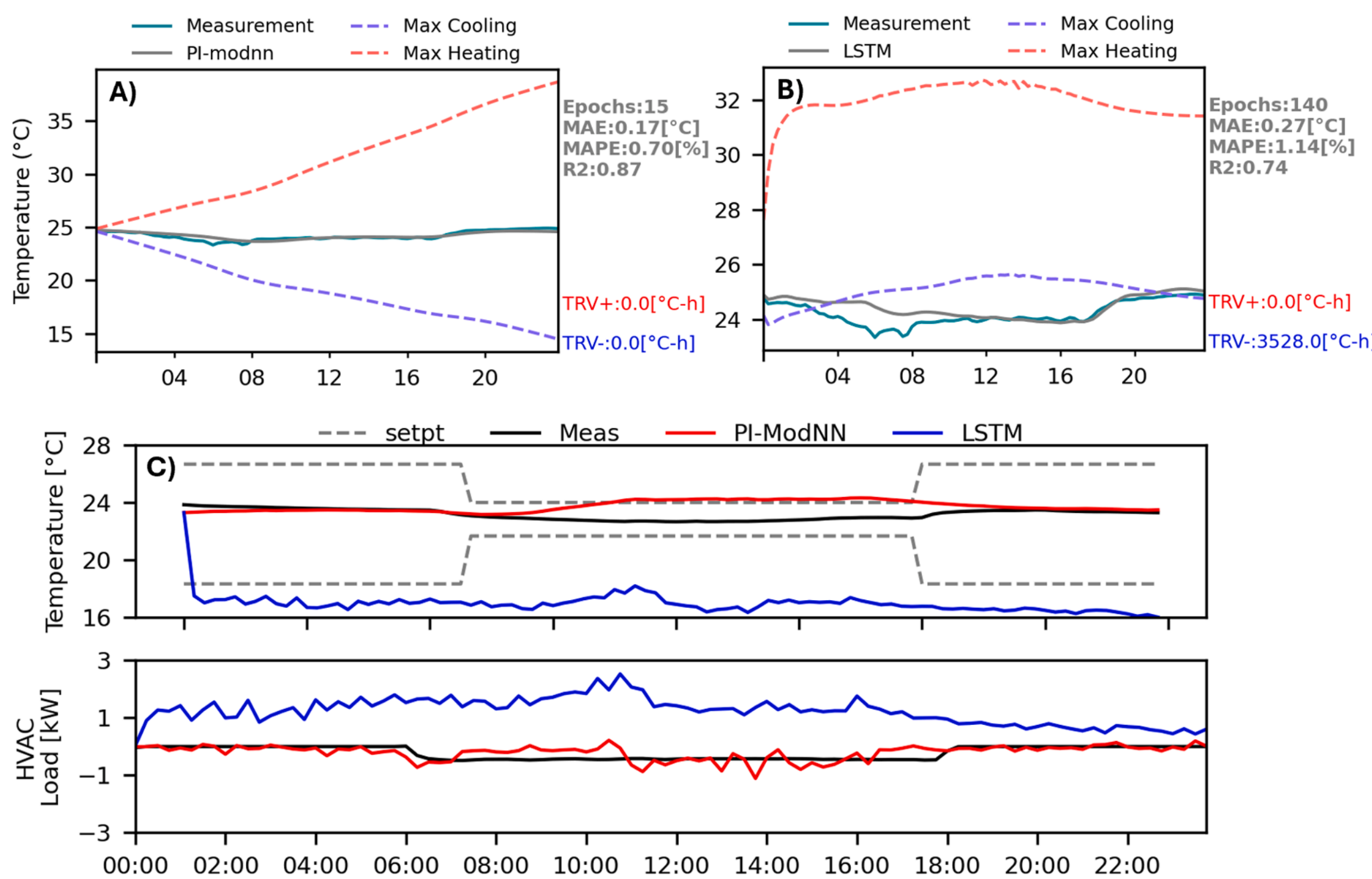

**Fig. A6.** A) Sanity check performance of the PI-ModNN environment, demonstrating physically consistent responses. B) Corresponding performance of the LSTM environment, which fails basic sanity checks—resulting in 3528 °C·h of temperature violation despite maximum cooling—and exhibits unrealistic behavior. C) Comparison of control performance using PI-ModNN (blue) and LSTM (red) environments on a representative summer day.

7. Over Estimated Space Air Temperature Due to Internal Heat Source Change

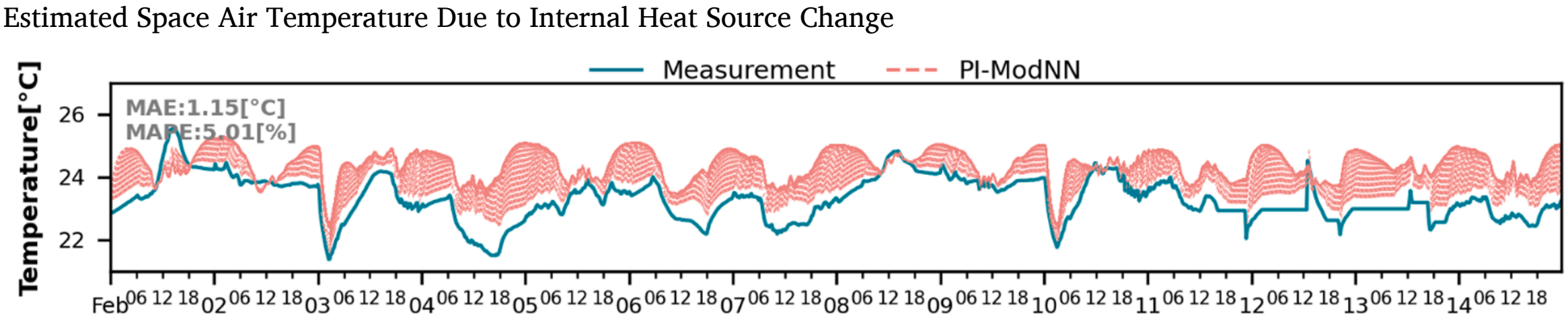

**Fig. A7.** Over-estimated space air temperature due to the changed internal heat gain.

8. Zero Temperature Violation of LSTM models

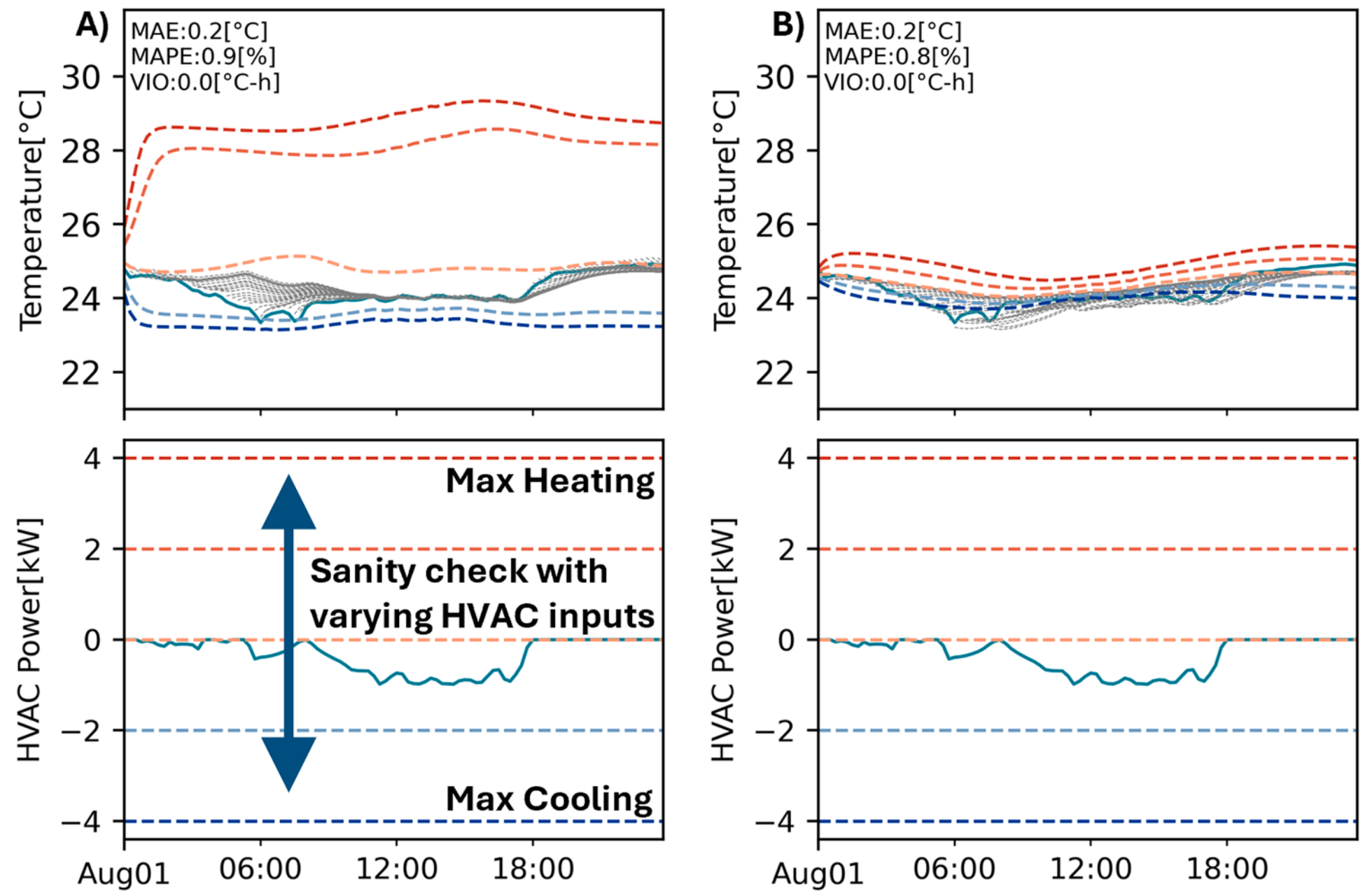

**Fig. A8.** LSTM achieves zero temperature response violation after multiple trials.

9. Training Loss Comparison

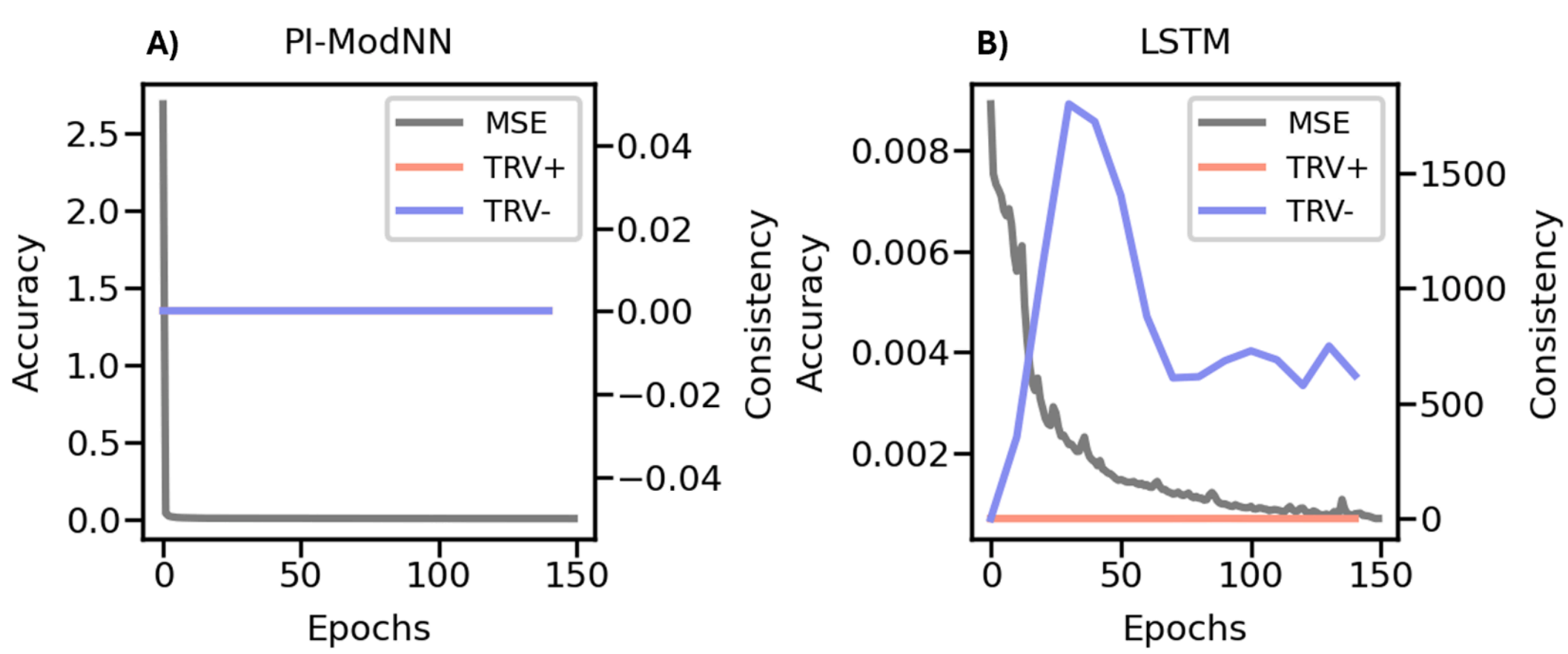

**Fig. A9.** Training loss curves. The accuracy loss is shown in gray, while the consistency losses are shown in red and blue, respectively. A) PI-ModNN model. B) LSTM model.

10. Code Availability

The proposed PI-ModNN has been open-sourced on GitHub: https://github.com/Bugs-Owner/Modularized-Neural-Network-Incorporating-Physical-Priors-for-Future-Building-Energy-Modeling, with a pip-installable package.

A Jupyter Notebook tutorial is also provided for demonstration and usage instructions: https://colab.research.google.com/drive/1A2jt1q53RtxGuaoym6N1PmlKELDPpYFX?usp=sharing.

## Data availability

Data will be made available on request.